\documentclass[11pt,draftclsnofoot,onecolumn]{IEEEtran}
\IEEEoverridecommandlockouts

\usepackage{epsfig}
\usepackage{amsthm}
\usepackage{amssymb}
\usepackage{setspace}

\usepackage{psfrag}
\usepackage{subfigure}
\usepackage{url}
\usepackage{stfloats}
\usepackage{amsmath}
\usepackage{array}
\usepackage[version=3]{mhchem}
\usepackage{subfigure}
\usepackage{amssymb,amsmath}

\usepackage{amsfonts}

\usepackage{multirow}

\usepackage{dsfont}
\usepackage[latin1]{inputenc}
\usepackage{times}

\usepackage{multibib}

\newtheorem{thm}{Theorem}
\newtheorem{rem}{Remark}
\newtheorem{cor}{Corollary}
\newtheorem{lem}{Lemma}
\newtheorem{dfn}{Definition}

\def\mb{\mathbf}

\def\bs{\boldsymbol}

\def\ss#1{{\sf #1}}

\def\resc#1{\uppercase{#1}}

\def\rvec#1{\bs{\uppercase{#1}}} 
\def\rvecr#1{\bs{\lowercase{#1}}} 

\def\mat#1{\mb{\uppercase{#1}}}
\def\mats#1{\bs{#1}}

\def\C{\ds{C}}

\def\EspOp{\ss{E}}

\newcommand{\Esp}[2][5]{%
  \ifcase#1
     \EspOp\{ #2 \}
     \or \EspOp \bigl\{ #2 \bigr\}
     \or \EspOp \Bigl\{ #2 \Bigr\}
     \or \EspOp \biggl\{ #2 \biggr\}
     \or \EspOp \Biggl\{ #2 \Biggr\}
  \else
     \EspOp \left\{ #2  \right\}
\fi}

\newcommand{\Earg}[3][5]{%
  \ifcase#1
     \EspOp_{#3} \{ #2 \}
     \or \EspOp_{#3} \bigl\{ #2 \bigr\}
     \or \EspOp_{#3} \Bigl\{ #2 \Bigr\}
     \or \EspOp_{#3} \biggl\{ #2 \biggr\}
     \or \EspOp_{#3} \Biggl\{ #2 \Biggr\}
  \else
     \EspOp_{#3} \left\{ #2  \right\}
\fi}

\newcommand{\CEsp}[3][5]{%
  \ifcase#1
     \EspOp\{ #2 \mid #3 \}
     \or \EspOp \bigl\{ #2 \bigm\vert #3 \bigr\}
     \or \EspOp \Bigl\{ #2 \Bigm\vert #3 \Bigr\}
     \or \EspOp \biggl\{ #2 \biggm\vert #3 \biggr\}
     \or \EspOp \Biggl\{ #2 \Biggm\vert #3 \Biggr\}
  \else
     \EspOp \left\{ #2  \,\middle\vert\, #3 \right\}
\fi}

\newcommand{\Diag}[2][5]{%
  \ifcase#1
     \mb{Diag}( #2 )
     \or \mb{Diag} \bigl( #2 \bigr)
     \or \mb{Diag} \Bigl( #2 \Bigr)
     \or \mb{Diag} \biggl( #2 \biggr)
     \or \mb{Diag} \Biggl( #2 \Biggr)
  \else
     \mb{Diag} \left( #2  \right)
\fi}

\newcommand{\diag}[2][5]{%
  \ifcase#1
     \mb{diag}( #2 )
     \or \mb{diag} \bigl( #2 \bigr)
     \or \mb{diag} \Bigl( #2 \Bigr)
     \or \mb{diag} \biggl( #2 \biggr)
     \or \mb{diag} \Biggl( #2 \Biggr)
  \else
     \mb{diag} \left( #2  \right)
\fi}

\def\Jacob{{\ss D}}

\def\Tr{\ss{Tr}}

\def\log{\ss{log}}

\def\d{\opn{d}\!}

\def\T{\ss{T}}

\def\ie{i.e.}
\def\eg{e.g.}
\def\iid{i.i.d.}

\def\EM{\mat{E}}
\def\MSE#1{\EM_{#1}}

\def\dim{n}
\def\I{ {I}} 

\def\scalart{\gamma}

\def\q{q}

\def\CM{\mats{\Phi}}
\def\EM{\mat{E}}
\def\MSE#1{\EM_{#1}}

\def\CMSEr#1#2{\CM_{#1}(#2)}

\def\Igen#1#2{ \I \left(  #1;#2 \right)}

\def\Icond#1#2#3{ \I \left( #1;#2 | #3 \right)}

\def\rate{\ss{R}}

\def\MSEcode{ \ss{MMSE}^{(C,\textit{f})}}
\def\MSEcoden{ \ss{MMSE}^{(C_n,\textit{f}_n)}}
\def\MMSE{ \ss{MMSE}}

\def\Hen#1{ \ss{H} \left( #1  \right)}
\def\Hbin#1{ \ss{H}_2 \left( #1  \right)}
\def\HenC#1#2{ \ss{H} \left( #1  | #2 \right)}
\def\d{ \ss{d}}
\def\C{ \ss{C}}

\def\snr{\ss{snr}}

\begin{document}

\title{On MMSE Properties of Codes for the Gaussian Broadcast and Wiretap Channels}
\author{\IEEEauthorblockN{Ronit Bustin, Rafael F. Schaefer, H. Vincent Poor and Shlomo Shamai (Shitz)\thanks{The work of R. Bustin was supported in part by the women postdoctoral scholarship of Israel's Council for Higher Education (VATAT) 2014-2015, in part by the the U. S. Army Research Office under MURI Grant W911NF-11-1-0036, and in part by the U. S. National Science Foundation under Grant ECCS-1343210. The work of R. F. Schaefer was supported by the German Research Foundation (DFG) under grant WY 151/2-1. The work of S. Shamai was supported by The Israeli Science Foundation (ISF) and the European FP7 Network of Excellence in Wireless COMmunications NEWCOM$\#$.}}
}

\maketitle


\begin{abstract}
This work concerns the behavior of ``good'' (capacity achieving) codes in several multi-user settings in the Gaussian regime, in terms of their minimum mean-square error (MMSE) behavior. The settings investigated in this context include the Gaussian wiretap channel, the Gaussian broadcast channel (BC) and the Gaussian BC with confidential messages (BCC). In particular this work addresses the effects of transmitting such codes on unintended receivers, that is, receivers that neither require reliable decoding of the transmitted messages nor are they eavesdroppers that must be kept ignorant, to some extent, of the transmitted message. This work also examines the effect on the capacity region that occurs when we limit the allowed disturbance in terms of MMSE on some unintended receiver. This trade-off between the capacity region and the disturbance constraint is given explicitly for the Gaussian BC and the secrecy capacity region of the Gaussian BCC.
\end{abstract}

\section{Introduction}
In this work we consider the scalar Gaussian channel with various requirements on transmission. The scalar Gaussian channel is a \emph{degraded} channel, meaning that a receiver at a higher signal-to-noise ratio (SNR) is a stronger receiver and can obtain all the information that a weaker receiver ($\ie$, a receiver at a lower SNR) can. The basic requirement is that of reliable communication. This requirement defines the Gaussian point-to-point channel, the capacity of which has been derived by Shannon in his seminal paper \cite{Shannon1948}. In \cite{EXIT} and \cite{StatisticalPhysics} the behavior of the input-output mutual information, at every SNR, assuming the input is a codeword from a ``good'' codebook sequence (a capacity achieving code sequence), has been investigated. It was shown that in the limit, as blocklength $\dim$ goes to infinity, this quantity follows the behavior of an independent and identically distributed ($\iid$) Gaussian input to the channel, up to the SNR at which reliable decoding is possible. This property is also related to \cite{ResolvabilityHanVerdu} where it was shown that the output statistics of a ``good'' code sequence approaches the output distribution produced by the $\iid$ capacity achieving input distribution. It follows that the behavior of the minimum mean-square error (MMSE) when estimating a codeword from a ``good'' codebook sequence from the channel output at different SNRs follows the MMSE of an $\iid$ Gaussian input to the channel \cite{EXIT,StatisticalPhysics}. As such the effect of a ``good'' code on receivers at lower SNRs in terms of the MMSE is well defined. More recently the effect of ``bad'' code sequences has been investigated \cite{IMMSEtradeoff}, that is, reliable codes of lower rates than the capacity. More specifically, the exact minimal MMSE that can be obtained by a code of rate $\rate$ was derived, and was shown to be that obtained by a superposition code sequence. This gives some engineering intuition into the efficiency of the Han-Kobayashi achievable region to the two-user Gaussian interference channel \cite{HanKobayashi}.

When considering potential receivers of the transmission at lower SNRs, an immediate requirement is \emph{confidentiality}. This is exactly the \emph{degraded} Gaussian wiretap channel, where we have one legitimate receiver, which requires reliable communication, and one eavesdropper, which must be kept as ignorant as possible with respect to the transmitted message. The \emph{degraded} wiretap channel was presented by Wyner \cite{WynerWireTap}, who showed that secrecy can be obtained simply by taking advantage of the physical layer - the channel to the legitimate receiver as compared to the channel of the eavesdropper. Since this initial work many have investigated the model and important results have been obtained: specifying the results for the Gaussian model \cite{WireTapGaussian}, extending to the general \emph{non-degraded} channel \cite{CsiszarKorner}, extensions to the Gaussian multiple-input/multiple-output (MIMO) channel, independently in \cite{KhistiWornell_journalPartI,KhistiWornell_journalPartII} and in \cite{OggierHassibi_journal} (for a more detailed review of the topic see \cite{LianPoorSham} and \cite{BlochBarros_PhysicalLayerScuirity} and references therein). Recently, more attention has been given to the design of ``good'' codes for the wiretap channel. This was first done for the discrete, \emph{degraded}, wiretap channel \cite{DiscreteWiretapCodes1,DiscreteWiretapCodes2,PolarCodeWiretap}. Tyagi and Vardi have shown a construction for the scalar Gaussian wiretap channel \cite{GaussianWiretapCode}. The approach in both cases was to use existing capacity approaching codes for the point-to-point channel to construct from them codes that comply with the secrecy constraint. More recently, Chou and Bloch considered the discrete broadcast channel (BC) with confidential messages and developed a low-complexity polar code scheme \cite{BlochPolarCodesBC}.

In this work we continue to examine the scalar Gaussian wiretap channel and specifically the properties of the MMSE function of specific families of codes for this setting. This analysis proves known ``rules of thumb'' in the design of such codes, and extends upon them to provide a good understanding to what is expected from a ``good'' code for this channel. Moreover, similar to the analysis of optimal point-to-point codes done in \cite{EXIT} and \cite{StatisticalPhysics}, such an understanding allows us to determine the effect/disturbance of these codes on other possible unintended receivers that are neither eavesdroppers nor require the reliable decoding of the message.

A different requirement to consider is reliable communication to an additional, weaker, receiver on top of the reliable communication to the stronger receiver. This defines the \emph{degraded} Gaussian BC. The BC was originally presented by Cover \cite{Cover_originalBC}, who also defined the \emph{degraded} BC, and conjectured its capacity region. The achievability proof of the \emph{degraded} BC was presented by Bergmans \cite{Bergmans_Achievble}, while the converse proof was derived by Gallager \cite{Gallager_Converse}. A converse proof for the Gaussian BC was derived  in parallel by Bergmans \cite{Bergmans_Converse} using the entropy power inequality (EPI). An alternative proof based on the single-letter expression and using the I-MMSE relationship was presented in \cite{PROP_full}. A comprehensive survey of the BC can be found in \cite{Comments} or \cite{CoverThomas}.

In \cite{StatisticalPhysics} the behavior of the MMSE function of a typical code sequence from a hierarchical code ensemble, designed to achieve capacity of the Gaussian BC, has been derived, via statistical physics methods. In this work we extend upon the above results and show that any ``good'' code sequence for the \emph{degraded} Gaussian BC exhibits the same behavior for the mutual information and MMSE functions. This behavior is that of a capacity achieving superposition code sequence, but holds for any ``good'' code sequence, that is, also code sequences designed using the ``Dirty Paper Coding'' (DPC) approach \cite{GelfandPinsker,WritingOnDirtyPaper}. As such, the full effect of ``good'' code sequences in term of the mutual information and MMSE functions is well defined. Moreover, we provide necessary and sufficient conditions for reliable decoding in general and for ``good'' code sequences for the \emph{degraded} scalar Gaussian BC, specifically. These conditions are given in the form of properties on the MMSE and conditional MMSE functions.

The two requirements mentioned above, that is, confidentiality and an additional weaker receiver, can be combined into a \emph{degraded} Gaussian BC with confidential messages (BCC). This problem has been considered already in the work of Csisz$\acute{\textrm{a}}$r and K$\ddot{\textrm{o}}$rner \cite{CsiszarKorner}, as the reliable rate to the weaker receiver can be considered as a rate of a common message (due to the \emph{degraded} nature of the channel). Csisz$\acute{\textrm{a}}$r and K$\ddot{\textrm{o}}$rner \cite{CsiszarKorner} provide a full single-letter solution for the rate equivocation region of this problem.

In this work we explore the case of maximum level of equivocation (secrecy) for the message sent to the stronger receiver in the BCC. We distinguish between the cases of maximum possible rate for this message and the case of complete secrecy. We show that in both cases the MMSE when estimating the codeword from the output follows that of the Gaussian BC, and the difference in the capacity region is due to the different behavior of the other relevant MMSE functions.

So far we have mentioned two types of requirements: reliable communication and confidentiality (equivocation). Another requirement concerns unintended receivers that are not interested in the transmission ($\ie$, do not require reliable communication) and are also not potential eavesdroppers ($\ie$, so no confidentiality is required). As shown in \cite{IMMSEtradeoff} an MMSE constraint at such receivers provides useful insights into the disturbance that the transmission imposes on these types of receivers. With respect to this we examine two problems: the capacity region of the \emph{degraded} Gaussian BC given an MMSE constraint and the secrecy capacity region of the \emph{degraded} Gaussian BCC given an MMSE constraint. In both problems we need to distinguish between two cases depending on the SNR of the unintended receiver. We show that the capacity achieving scheme in both is based on superposition, whereas in the Gaussian BCC, depending on the SNR of this receiver, we can obtain capacity by using either a superposition code sequence for the wiretap code sequence to the stronger receiver, or a superposition code sequence to the weaker receiver.


As this work concerns the MMSE performance of code sequences, we use the fundamental relationship between information theory and estimation theory, namely the so called I-MMSE relationship \cite{IMMSE}. As we discuss ``good'' code sequences, we consider the behavior of the MMSE function in the limit as $\dim \to \infty$. This limit might not always exist; however if we restrict ourselves to information stable distributions (for which the information rate is defined), we can discuss the $\limsup$ of the MMSE function. We show that the relevant properties used in the context of the limit can also be considered with regard to the $\limsup$. Specifically, this holds for the ``single crossing point'' property, originally derived in \cite{PROP_full}, which examines the difference between the MMSE function assuming a Gaussian input distribution and the MMSE function of any arbitrary input distribution. This property shows that this difference has at most a single crossing of the horizontal axis, meaning that if for some SNR the two MMSE functions are equal, for any higher SNR the MMSE function of the Gaussian input will attain a higher value (assuming the two are not equal for all SNR, which occurs only when both have the same Gaussian input distribution). In this paper we extend the simplest vector version of this property, derived in \cite{Full_BustinPayaroPalomarShamai}, to a conditioned version.



\underline{Notation:} We denote a random variable using uppercase letter, such as $\resc{x}$. Its realizations are denoted using a lower case letter ($\eg$, $x$). We denote a length $\dim$ random vector with bold uppercase letter and the subscript $\dim$ ($\eg$, $\rvec{x}_{\dim}$). Specific realization of the random vector are denoted using the equivalent lower case ($\eg$, $\rvecr{x}_{\dim}$). When we consider quantities after taking the dimension to infinity, $\dim \to \infty$, we remove all subscripts $\dim$. The trace operator is denoted by $\Tr$ and the transpose operator is denoted by $( \cdot )^\T$. The standard vector norm is denoted by $\parallel \cdot \parallel$. The expectation operator is denoted by $\EspOp$ and the derivative of some function with respect to one of its parameter, for example $\gamma$, is denoted by $\Jacob_{\gamma}$.

The paper is structured as follows: we begin with some preliminary results and definitions in Section \ref{sec:Preliminary}. The Gaussian wiretap problem is discussed in Section \ref{sec:Wiretap}. The Gaussian BC problem is discussed in Section \ref{sec:GaussianBC}. The Gaussian BCC is discussed in Section \ref{sec:GaussianBCC}, and the disturbance constraints are considered in Section \ref{sec:Disturbance}. We then conclude and summarize the paper.





\section{Preliminary Definitions} \label{sec:Preliminary}
In this section we provide the central definitions and results that will be used throughout the paper.

\subsection{The Code Sequences} \label{ssec:codesequences}
In this work we split the definition of a code sequence and distinguish between the set of codewords and the mapping of messages to these codewords.
We define a codebook sequence $\C = \left\{ \C_{\dim} \right\}_{\dim=1}^{\infty}$ where $\C_{\dim}$ is a set of length-$\dim$ codewords denoted as the random vector $\rvec{x}_{\dim}$ with realizations denoted as $\rvecr{x}_{\dim}$. These codewords must comply with a power constraint. With no loss of generality we assume the following:
\begin{align}
\frac{1}{\dim} \parallel \rvecr{x}_{\dim} \parallel^2 \leq 1, \quad \forall \rvecr{x}_{\dim} \in \C_{\dim}.
\end{align}
We further define a mapping sequence $f = \left\{f_{\dim} \right\}_{\dim = 1}^{\infty}$ where $f_{\dim}$ is a mapping from a message $W$ to a codeword of length-$\dim$, $\ie$,
\begin{align}
f_{\dim}(W) = \rvecr{x}_{\dim}.
\end{align}
The standard definition of a code sequence is a pair $(\C, f)$. Such a pair is said to have rate $\rate$ if the message $W$ is one of $\{1,2, \ldots, 2^{\dim \rate} \}$ (assumed uniformly distributed over the set) and is mapped to a codeword $\rvecr{x}_{\dim} \in \C_{\dim}$.
Note that these mappings are not necessarily deterministic,
specifically if confidentiality is required then stochastic encoding is necessary \cite[Section 3.4.1]{BlochBarros_PhysicalLayerScuirity},
in which case, for each message $W_i$ there is a subset $S_i \subset \C_{\dim}$ such that $f_{\dim}(W_i) \in S_i$. These subsets are disjoint in order to allow reliable communication over the channel.

The code sequence pairs $(\C, f)$ considered in this paper are assumed to be information stable, meaning all mutual information rates are assumed to converge as $\dim \to \infty$.

\subsection{The I-MMSE Approach} \label{ssec:IMMSEapproach}
The main approach used here is the I-MMSE approach, that is to say that we make use of the fundamental relationship between the mutual information and the MMSE in the Gaussian channel and its generalizations
\cite{IMMSE,Palomar,FnT_IMMSE,ShlomosLetter}. Even though we are examining scalar settings, the $\dim$-dimensional version of this relationship is required since we are looking at the transmission of length-$\dim$ codewords through the channel. In our setting the relationship is as follows:
\begin{eqnarray} \label{eq:IMMSErelationship2}
\frac{1}{\dim} \Igen{\rvec{x}_{\dim}}{\sqrt{\snr} \rvec{x}_{\dim} + \rvec{n}_{\dim}} = \frac{1}{2} \int_0^\snr \MSEcoden(\gamma) \d \gamma
\end{eqnarray}
where $\rvec{n}_{\dim}$ is standard additive Gaussian noise, $\rvec{x}_{\dim}$ is the input signal of any arbitrary distribution (as long as the above mutual information is finite \cite{FunctionalPropertiesMMSE}), $\MSEcoden(\gamma) = \frac{1}{\dim} \Tr( \MSE{\rvec{x}_{\dim}}(\gamma) )$ and $\MSE{\rvec{x}_{\dim}}(\gamma)$ is the MMSE matrix defined as follows:
\begin{align}
\MSE{\rvec{x}_{\dim}}(\gamma) = \EspOp \bigl\{ (\rvec{x}_{\dim} - \CEsp{\rvec{x}_{\dim}}{\sqrt{\gamma} \rvec{x}_{\dim} + \rvec{n}_{\dim}})  (\rvec{x}_{\dim} - \CEsp{\rvec{x}_{\dim}}{\sqrt{\gamma} \rvec{x}_{\dim} + \rvec{n}_{\dim}})^\T \bigr\}.
\end{align}
Note that in our setting the distribution of $\rvec{x}_{\dim}$ is defined by the mapping $f_{\dim}$ and the assumption of uniformly distributed messages, $W$.

The I-MMSE relationship has also been extended to conditional cases \cite{PROP_full}. Consider a joint distribution over $(\rvec{x}_{\dim}, \resc{u})$ independent of the additive Gaussian noise. Denote the random vector $\rvec{x}_{\dim, u}$ with distribution $P_{\rvec{x} | \resc{u} = u}$. Applying the I-MMSE relationship we have
\begin{align}
\frac{1}{\dim}\Igen{\rvec{x}_{\dim,u}}{\sqrt{\snr} \rvec{x}_{\dim,u} + \rvec{n}_{\dim}} = \frac{1}{2} \int_{0}^{\snr} \MSEcoden( \gamma | \resc{u} = u ) \d \gamma \label{eq:IMMSEconditioned1}
\end{align}
where $\MSEcoden(\gamma | \resc{u} = u) = \frac{1}{\dim} \Tr( \MSE{\rvec{x}_{\dim}}(\gamma, u) )$ and
\begin{align}
\MSE{\rvec{x}_{\dim}}(\gamma, u) = \EspOp \bigl\{ (\rvec{x}_{\dim,u} - \CEsp{\rvec{x}_{\dim,u}}{\sqrt{\gamma} \rvec{x}_{\dim,u} + \rvec{n}_{\dim}}) (\rvec{x}_{\dim,u} - \CEsp{\rvec{x}_{\dim,u}}{\sqrt{\gamma} \rvec{x}_{\dim,u} + \rvec{n}_{\dim}})^\T \bigr\}.
\end{align}
Since $(\rvec{x}_{\dim}, \resc{u})$ is independent of the noise we have that
\begin{multline}
\MSE{\rvec{x}_{\dim}}(\gamma, u) = \EspOp \bigl\{ (\rvec{x}_{\dim} - \CEsp{\rvec{x}_{\dim}}{\sqrt{\gamma} \rvec{x}_{\dim} + \rvec{n}_{\dim}, \resc{u} = u}) \bigr. \\
\bigl. (\rvec{x}_{\dim} - \CEsp{\rvec{x}_{\dim}}{\sqrt{\gamma} \rvec{x}_{\dim} + \rvec{n}_{\dim}, \resc{u} = u})^\T | \resc{u} = u\bigr\}.
\end{multline}
Taking the expectation over $\resc{u}$ results with
\begin{align}
\Esp{ \MSE{\rvec{x}_{\dim}}(\gamma, \resc{u}) } = \EspOp \bigl\{ (\rvec{x}_{\dim} - \CEsp{\rvec{x}_{\dim}}{\sqrt{\gamma} \rvec{x}_{\dim} + \rvec{n}_{\dim}, \resc{u} }) (\rvec{x}_{\dim} - \CEsp{\rvec{x}_{\dim}}{\sqrt{\gamma} \rvec{x}_{\dim} + \rvec{n}_{\dim}, \resc{u} })^\T \bigr\}
\end{align}
and $\MSEcoden(\gamma | \resc{u} ) = \Esp{ \frac{1}{\dim} \Tr( \MSE{\rvec{x}_{\dim}}(\gamma, \resc{u}) )}$.
Moreover, this independence also allows us to write (\ref{eq:IMMSEconditioned1}) as
\begin{align}
\frac{1}{\dim}\Icond{\rvec{x}_{\dim}}{\sqrt{\snr} \rvec{x}_{\dim} + \rvec{n}_{\dim}}{\resc{u} = u} = \frac{1}{2} \int_{0}^{\snr} \MSEcoden( \gamma | \resc{u} = u ) \d \gamma. \label{eq:IMMSEconditioned}
\end{align}
Taking the expectation on both sides of (\ref{eq:IMMSEconditioned}) with respect to $\resc{u}$ we obtain the following conditioned version of the I-MMSE relationship:
\begin{align}
\frac{1}{\dim}\Icond{\rvec{x}_{\dim}}{\sqrt{\snr} \rvec{x}_{\dim} + \rvec{n}_{\dim}}{\resc{u}} & = \frac{1}{2} \int_{0}^{\snr} \MSEcoden( \gamma | \resc{u}) \d \gamma.
\end{align}

We now take the limit as $\dim \to \infty$ on both sides of the I-MMSE relationship (\ref{eq:IMMSErelationship2}). In order to simplify notation the quantities at the limit, as $\dim \to \infty$ are simply denoted by removing the subscript $\dim$. For code sequences for which the relevant MMSE quantities converge as $\dim \to \infty$, the exchange of limit and integration is according to Lebesgue's dominated convergence theorem \cite{Lebesgue}, since the MMSE quantities are always bounded. Thus, we have
\begin{align} \label{eq:IMMSErelationship3_infty}
\Igen{\rvec{x}}{\sqrt{\snr} \rvec{x} + \rvec{n}} & \equiv \lim_{\dim \to \infty} \frac{1}{\dim} \Igen{\rvec{x}_{\dim}}{\sqrt{\snr} \rvec{x}_{\dim} + \rvec{n}_{\dim}} \nonumber \\
& = \lim_{\dim \to \infty} \frac{1}{2} \int_0^\snr \MSEcoden(\gamma) \d \gamma \nonumber \\
 & = \frac{1}{2} \int_0^\snr \MSEcode(\gamma) \d \gamma .
\end{align}
When this is not the case ($\ie$, the MMSE does not converge) we can apply the reverse Fatou's Lemma \cite{Lebesgue} and conclude that
\begin{align} \label{eq:ReverseFatou}
\lim_{\dim \to \infty} \int_0^{\snr} \MSEcoden(\gamma) \d \gamma = \int_0^{\snr} \MSEcode(\gamma)_{sup} \d \gamma
\end{align}
where
\begin{align} \label{eq:defineMMSEsup}
\MSEcode(\gamma)_{sup} = \limsup_{\dim \to \infty} \MSEcoden(\gamma)
\end{align}
due to the stability of the information rates. We show this precisely in Appendix \ref{appendix:Fatou}.
In this case the MMSE quantities throughout the paper are the $\limsup$ of the MMSE sequences, which always exist and, since the MMSE function is bounded, are also finite.

Another property used in the proof is a simple extension of the  $\dim$-dimensional ``single crossing point'' property derived in \cite{Full_BustinPayaroPalomarShamai} to the conditioned case considered here.
For an arbitrary random vector $\rvec{x}_{\dim}$ we consider the following function
\begin{IEEEeqnarray}{rCl}
\q(\rvec{x}_{\dim} | \resc{u}, \sigma^2, \scalart) & = & \frac{\sigma^2}{1
+ \sigma^2 \scalart}  - \Esp{ \frac{1}{\dim} \Tr \left(\MSE{\rvec{x}_{\dim}}(\scalart, \resc{u} ) \right) } \\
& = &\frac{\sigma^2}{1 + \sigma^2 \scalart} - \MSEcoden(\gamma | \resc{u}) . \label{eq:defqA}
\end{IEEEeqnarray}
Note that $\frac{\sigma^2}{1 + \sigma^2 \scalart}$ is the MMSE function assuming the input distribution is $\iid$ Gaussian with variance $\sigma^2$. Thus, the above function is the difference between the MMSE function of an $\iid$ Gaussian input and the averaged MMSE function assuming an arbitrary joint distribution over $(\rvec{x}_{\dim}, \resc{u})$, that is independent of the Gaussian noise.
The following theorem is proved in Appendix \ref{appendix:SingleCrossingPointExt}:
\begin{thm} \label{thm:ScalarUniqueCrossingPoint}
The function $\scalart \mapsto \q(\rvec{x}_{\dim}|\resc{u}, \sigma^2, \scalart)$, defined
in (\ref{eq:defqA}), has no nonnegative-to-negative zero crossings and, at most,
a single negative-to-nonnegative zero crossing in the range $\scalart \in [0,
\infty)$. Moreover, let $\snr_0 \in [0,  \infty)$ be that
negative-to-nonnegative crossing point. Then,
\begin{enumerate}
\item $\q(\rvec{x}_{\dim} | \resc{u}, \sigma^2, 0) \leq 0$.
\item $\q(\rvec{x}_{\dim}| \resc{u}, \sigma^2, \scalart)$ is a strictly increasing
function in the range $\scalart \in [0, \snr_0)$.
\item $\q(\rvec{x}_{\dim}| \resc{u}, \sigma^2, \scalart) \geq 0$ for all $\scalart \in
[\snr_0, \infty)$.
\item $\lim_{\scalart \to \infty} \q(\rvec{x}_{\dim}| \resc{u}, \sigma^2, \scalart) =
0$.
\end{enumerate}
\end{thm}
The above properties are valid for all natural $\dim$. As explained above, if the limit of $\MSEcoden(\gamma | \resc{u})$ exists then we may simply take the limit of the function $\q( \rvec{x}_{\dim}| \resc{u}, \sigma^2, \scalart)$. 
Alternatively, we show in Appendix \ref{appendix:singleCrossing} that the ``single crossing point'' property holds also between $\frac{\sigma^2}{1 + \sigma^2 \scalart}$ and $\limsup \MSEcoden(\gamma| \resc{u})$.

Finally, in this section we denoted the MMSE function using $\MSEcoden(\gamma)$ (and its conditional version as $\MSEcoden(\gamma | \resc{u})$) so as to emphasize the dependence on the code sequence. In the sequel we will simplify the notation and use the following:
\begin{align}
\MMSE ( \rvec{x}_{\dim} ; \gamma ) & = \MSEcoden(\gamma)
\end{align}
and $\MMSE ( \rvec{x}_{\dim} ; \gamma | \resc{U})$ for the conditioned version. 

\section{The Gaussian Wiretap Channel} \label{sec:Wiretap}

\subsection{Model and Definitions} \label{ssec:wiretapModelDef}
The Gaussian wiretap channel \cite{WynerWireTap}, over which length-$\dim$ codewords are being transmitted, is denoted here as follows:
\begin{align}
\rvec{y}_{\dim} = \sqrt{ \snr_y } \rvec{x}_{\dim} + \rvec{n_1}_{\dim} \nonumber \\
\rvec{z}_{\dim} = \sqrt{ \snr_z } \rvec{x}_{\dim} + \rvec{n_2}_{\dim}
\end{align}
where $\rvec{x}_{\dim}$ is the length-$\dim$ transmitted codeword. $\rvec{n_1}_{\dim}$ and $\rvec{n_2}_{\dim}$ are standard additive Gaussian noise vectors that can be assumed independent of each other, and $\snr_y$ and $\snr_z$ are the SNRs at the two receivers. We assume that $\snr_y > \snr_z$. $\rvec{y}$ represents the legitimate receiver, and $\rvec{z}$ represents the eavesdropper. A message $W_y$ is transmitted to $\rvec{y}$ and needs to be reliably decoded. The message must be kept secret, to some extent, from the eavesdropper.


An $(\rate, d)$ code sequence for this channel must reliably transmit the message, $W_y$, of cardinality $2^{\dim \rate}$, to the legitimate receiver, and also guarantee an equivocation rate of $d$. These requirements can be written as follows:
\begin{align} \label{eq:wiretapDefinition}
\lim_{\dim \to \infty} \frac{1}{\dim} \Igen{W_y}{\rvec{y}_{\dim}} & = \rate \nonumber \\ 
\lim_{\dim \to \infty} \frac{1}{\dim} H(W_y | \rvec{z}_{\dim}) & = d.
\end{align}
The basic requirement from the pair $(\rate, d)$ is that $\rate \geq d$. 
As stated in \cite{CsiszarKorner} these conditions do not specify the joint conditional distribution of the outputs given the channel inputs, but only the marginals of this joint conditional distribution have an effect. As such, we may assume, without loss of generality, $\rvec{n_1}_{\dim} = \rvec{n_2}_{\dim} = \rvec{n}_{\dim}$.
The rate-equivocation region $(\rate, d)$  for this model was obtained by Wyner \cite{WynerWireTap} and is as follows:
\begin{align} \label{eq:RateEquivocation}
C = \left\{ \begin{array}{l} (\rate, d): \\ 0 \leq \rate \leq \Igen{\resc{x}}{\resc{y}} \\
0 \leq d \leq \rate \\
d \leq  \left[ \Igen{\resc{x}}{\resc{y}} - \Igen{\resc{x}}{\resc{z}} \right] \end{array} \right..
\end{align}
In the Gaussian setting considered here the capacity region is well-known \cite{WireTapGaussian}, and is obtained by the standard Gaussian input in the above expression.
We wish to distinguish between three rate-equivocation phenomena, which we define next.
\begin{dfn} \label{dfn:perfectSecrecy}
A code sequence obtains \emph{complete secrecy} when $\rate = d$\footnote{\emph{Complete secrecy} is sometimes referred to in the literature as \emph{full secrecy} \cite{BlochBarros_PhysicalLayerScuirity} or \emph{perfect secrecy}.}.
\end{dfn}
\begin{dfn} \label{dfn:optimallySecure}
A code sequence is \emph{optimally secure} when
\begin{align}
d = d_{opt} = \lim_{\dim \to \infty} \frac{1}{\dim} \left[ \Igen{\rvec{x}_{\dim}}{\rvec{y}_{\dim}} - \Igen{\rvec{x}_{\dim}}{\rvec{z}_{\dim}} \right].
\end{align}
\end{dfn}
\begin{dfn} \label{dfn:maximumRate}
A code sequence $(\rate, d)$ is at its \emph{maximum possible rate} if
\begin{align}
\rate = \rate_{\max} = \lim_{\dim \to \infty} \frac{1}{\dim} \Igen{\rvec{x}_{\dim}}{\rvec{y}_{\dim}}.
\end{align}
\end{dfn}
We further define two quantities. The \emph{maximum level of equivocation},
\begin{align} \label{def:MAXequivocation}
d_{\max} = \frac{1}{2} \log(1 + \snr_y) - \frac{1}{2} \log(1 + \snr_z);
\end{align}
and the \emph{capacity of the point-to-point channel}, which is also the maximum possible rate for the Gaussian wiretap channel, for any possible $d$,
\begin{align} \label{def:MAXrate}
C = \frac{1}{2} \log(1 + \snr_y).
\end{align}

\subsection{Main Results} \label{ssec:WiretapMainResults}
In this subsection we provide our main results and observations on code sequences for the scalar Gaussian wiretap channel. The proofs of these results are given in the next subsection.

\begin{thm} \label{thm:optimallySecureCode}
Any code sequence for the Gaussian wiretap channel attains $(\rate, d_{opt})$, that is, optimal security, if and only if
\begin{align} \label{thm:eq:MMSE_Wbehavior}
\MMSE( \rvec{x} ; \gamma | W_y) = 0, \quad \gamma \geq \snr_z
\end{align}
and
\begin{align} \label{thm:eq:MMSEbehavior}
\MMSE( \rvec{x}; \gamma) = 0, \quad \gamma \geq \snr_y
\end{align}
regardless of the rate of the code ($\rate \geq d_{opt}$).
\end{thm}
Note that \emph{optimally secure} codes do not necessarily contain any confidential information; for example, any ``good'' point-to-point code sequence to $\rvec{z}$ will also be \emph{optimally secure} with $d = d_{opt} = 0$ (as $\Igen{\rvec{x}}{\rvec{y}} = \Igen{\rvec{x}}{\rvec{z}}$). Nonetheless, the importance of the above result is in emphasizing two properties of the family of code sequences that are \emph{optimally secure}. First, at $\snr_y$ we can fully decode the transmitted codeword and not only the message sent. Second, given the message, full decoding of the codeword will be possible at and above $\snr_z$. This observation supports the capacity achieving schemes in which the knowledge of the message leads to full decoding of the codeword by the eavesdropper. Note that the above is valid for any point $(\rate, d_{opt})$ regardless of $\rate$ ($\rate \geq d_{opt}$).

The above behavior can be further specified for the particular case of $d_{opt} = d_{\max}$ as follows:
\begin{cor} \label{cor:d_maxObservations}
Any code sequence for the Gaussian wiretap channel attains $( \rate, d_{\max})$, that is, the maximum level of equivocation, if and only if
\begin{align} \label{cor:eq:MMSE_Wbehavior}
\MMSE \left( \rvec{x}; \gamma |  W_y \right) = 0, \quad \gamma \geq \snr_z
\end{align}
and
\begin{align} \label{cor:eq:MMSEbehavior}
\MMSE \left( \rvec{x}; \gamma \right) = \left\{ \begin{array}{ll} \frac{1}{1 + \gamma}, & \gamma \in [0, \snr_y) \\
0, & \gamma \geq \snr_y \end{array} \right.
\end{align}
regardless of the rate of the code ($\rate \geq d_{\max}$).
\end{cor}

Note that the behavior in (\ref{cor:eq:MMSEbehavior}) is the behavior of any ``good'' point-to-point code sequence, $\C$, to $\rvec{y}$ \cite{EXIT,StatisticalPhysics}; however, only a one-to-one mapping over this codebook sequence leads to a maximum point-to-point rate.
(\ref{cor:eq:MMSEbehavior}) also suggests that there are approximately $2^{\dim \frac{1}{2} \log( 1+ \snr_y)}$ codewords in any codebook in the sequence of maximum equivocation codebooks.
Note that this property defines a large family of code sequences which contains those for the Gaussian wiretap channel with $(\rate, d_{\max})$.
The idea is that the maximum level of equivocation determines $\MMSE( \rvec{x} ; \gamma )$ for every $\gamma$ regardless of the rate. In other words, in order to obtain a maximum level of equivocation the codewords must resemble the $\iid$ Gaussian distribution over the channel (in terms of MMSE) up to an SNR of $\snr_y$.

The additional condition given in (\ref{cor:eq:MMSE_Wbehavior}) (or alternatively, in (\ref{thm:eq:MMSE_Wbehavior})) is required in order to fully define the sub-group of code sequences that are $(\rate, d_{\max})$ (or alternatively, $(\rate, d_{opt})$) codes for the Gaussian wiretap channel. Still, these conditions do not fully specify the rate of the code sequence, as the group contains codes of different rates $\rate$ as long as $\rate \geq d_{\max}$ (or alternatively, $\rate \geq d_{opt}$).
The immediate question that arises is: Can we find MMSE properties that will distinguish code sequences of different rates?
The next theorem provides some insight into two families of code sequences, the complete secrecy, $\rate = d$, family and the maximum rate, $\rate = \rate_{\max}$, family.


\begin{thm} \label{thm:specificRates}
A code sequence for the Gaussian wiretap channel is an $(\rate, \rate)$ code, that is, it attains complete secrecy $(\rate = d)$, if and only if 
it has the following behavior:
\begin{align}
\label{thm:eq:perfectSecrecy}
\MMSE( \rvec{x}; \gamma | W_y) & = \MMSE( \rvec{x} ; \gamma), \quad \forall \gamma \in [0, \snr_z) \intertext{ and } \label{thm:eq:PerfectSecrecyReliableComm}
\rate & = \frac{1}{2} \int_{\snr_z}^{\snr_y} \MMSE( \rvec{x} ; \gamma ) - \MMSE( \rvec{x}; \gamma |  W_y) \d \gamma.
\end{align}
A code sequence for the Gaussian wiretap channel attains $\rate_{\max}$ if and only if 
it has the following behavior:
\begin{align} \label{thm:eq:MMSE_Wbehavior_secrecy}
\MMSE \left( \rvec{x}; \gamma | W_y \right) = 0, \quad \forall \gamma \geq 0,
\end{align}
meaning a one-to-one mapping of the messages to codewords. \\
Specifically, if $\rate = d_{\max}$ we also have the properties of Corollary \ref{cor:d_maxObservations}.
\end{thm}

\begin{figure}[h]
\begin{center}
\setlength{\unitlength}{.1cm}
    \includegraphics[width=1.1\textwidth]{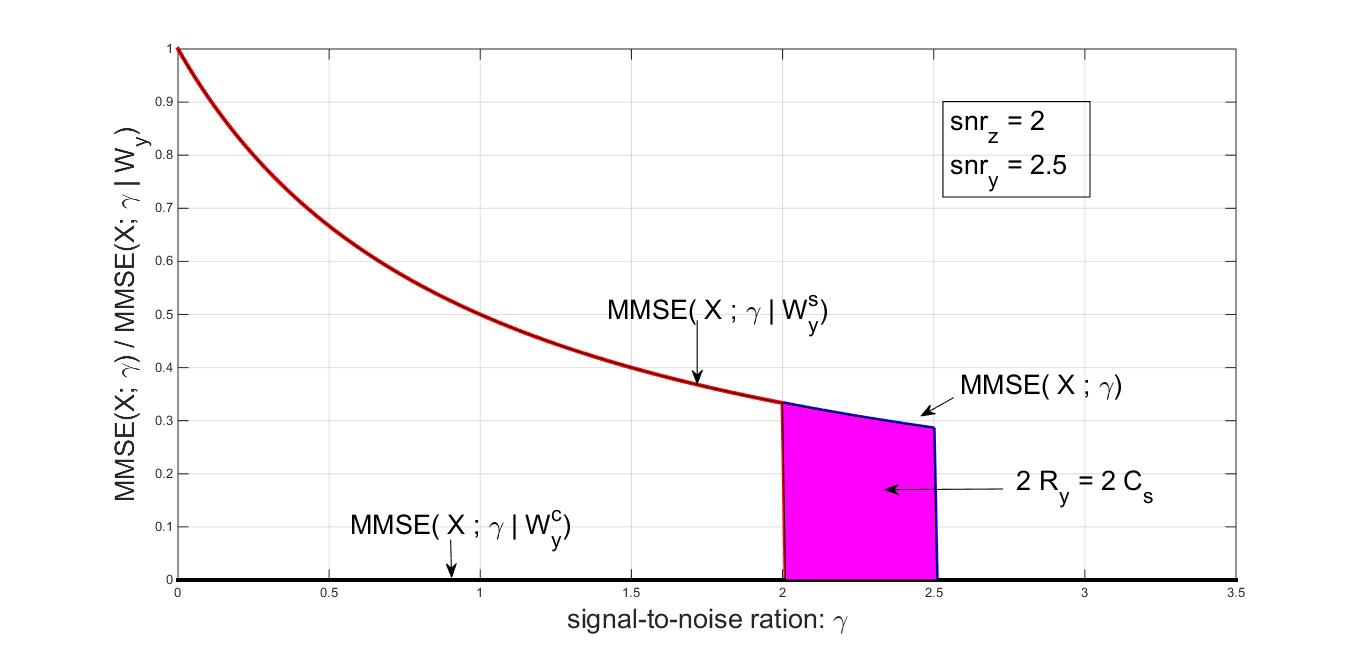}
   \caption{The above figure depicts the behavior of $\MMSE( \rvec{x}; \gamma)$ as a function of $\gamma$ assuming $d_{\max}$ (in blue), the behavior of $\MMSE(\rvec{x}; \gamma | W_y^s)$ assuming complete secrecy (in red) and the behavior of $\MMSE(\rvec{x}; \gamma | W_y^c)$ assuming a ``good'' point-to-point code (in black). Finally, we mark twice the secrecy capacity rate as the region between $\MMSE( \rvec{x}; \gamma)$ and $\MMSE(\rvec{x}; \gamma | W_y^s)$ (in magenta).}
 \label{figure:wiretap_perfectSecrecy}
\end{center}
\end{figure}

\begin{figure}[h]
\begin{center}
\setlength{\unitlength}{.1cm}
    \includegraphics[width=1.1\textwidth]{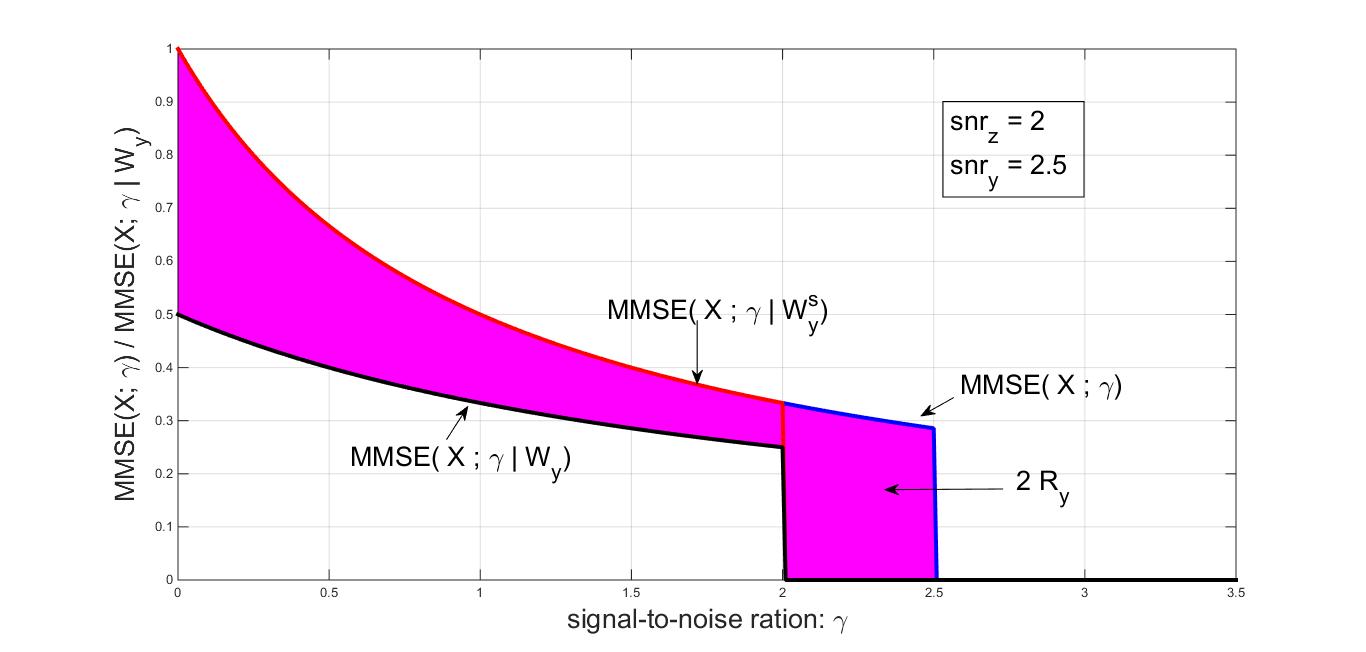}
    \caption{The above figure we depicts the behavior of $\MMSE( \rvec{x}; \gamma)$ as a function of $\gamma$ assuming $d_{\max}$ (in blue), the behavior $\MMSE(\rvec{x}; \gamma | W_y^s)$ assuming complete secrecy (in red) and the behavior of $\MMSE(\rvec{x}; \gamma | W_y)$ for some arbitrary code of rate above secrecy capacity and below point-to-point capacity (in black). We mark twice the rate as the area between $\MMSE( \rvec{x}; \gamma)$ and $\MMSE(\rvec{x}; \gamma | W_y)$ (in magenta). }
 \label{figure:wiretap_inter}
\end{center}
\end{figure}

From Corollary \ref{cor:d_maxObservations} and Theorem \ref{thm:specificRates} (see equation (\ref{eq:rateOfWiretap}) in the proof of Theorem \ref{thm:specificRates}) we can conclude that when $d = d_{\max}$ the rate of the code is determined solely by the behavior of $\MMSE \left( \rvec{x}; \gamma | W_y \right)$ in the region of $\gamma \in [0, \snr_z)$ (and is either equal to or above the secrecy capacity rate). Note that this is not the case when $d < d_{\max}$ since $\MMSE( \rvec{x} ; \gamma | W_y)$ does not necessarily fall to zero at $\snr_z$ (if $d < d_{opt}$ ), and the behavior of $\MMSE( \rvec{x} ; \gamma)$ is not predetermined.
As shown in Theorem \ref{thm:specificRates}, when we require only complete secrecy, the expression for the rate is given by (\ref{thm:eq:PerfectSecrecyReliableComm}), a function of both $\MMSE( \rvec{x} ; \gamma | W_y)$ and  $\MMSE( \rvec{x} ; \gamma )$ in the region of $[\snr_z, \snr_y)$.  
These observations are also depicted in Figures \ref{figure:wiretap_perfectSecrecy} and \ref{figure:wiretap_inter} where we depict the behavior assuming a $d_{\max}$ code sequence. In Figure \ref{figure:wiretap_perfectSecrecy} we consider a \emph{completely secure} code sequence, and in Figure \ref{figure:wiretap_inter} we consider the behavior once we increase the rate beyond the secrecy capacity.

Another important thing to observe is that as the condition (\ref{thm:eq:MMSE_Wbehavior_secrecy}) is a necessary and sufficient condition it supports the necessity of a stochastic encoder for any code sequence for the Gaussian wiretap channel with $\rate < \rate_{\max}$ (as shown in \cite[Section 3.4.1]{BlochBarros_PhysicalLayerScuirity} for a completely secure code for the discrete memoryless wiretap channel). This is due to the fact that
\begin{align}
\Hen{ \rvec{x} | W_y } \geq \Icond{ \rvec{x}}{ \sqrt{\snr} \rvec{x} + \rvec{n} }{W_y}
\end{align}
and
\begin{align}
\Icond{ \rvec{x}}{ \sqrt{\snr} \rvec{x} + \rvec{n} }{W_y} = \frac{1}{2} \int_0^{\snr} \MMSE( \rvec{x}; \gamma |W_y) \d \gamma.
\end{align}
Thus, condition (\ref{thm:eq:MMSE_Wbehavior_secrecy}) guarantees $\Hen{ \rvec{x} | W_y } > 0$ for any such code sequence.

Note that a central conclusion from Corollary \ref{cor:d_maxObservations} is that any maximum equivocation code sequence has a ``good'' point-to-point  codebook sequence. In such a case even when the rate is below capacity we can artificially complete the mapping of messages to codewords to a one-to-one mapping. This leads us to investigate the ``good'' point-to-point sequences. The following result basically restates the known property that such sequences are also $(C, d_{\max})$ achieving for the Gaussian wiretap channel.

\begin{lem} \label{lem:P2Poptimal2WiretapOptimal}
For any $\snr_z' \in [0, \snr_y)$ define
\begin{align} \label{eq:thm:P2Poptiaml_dmax_tag}
d_{\max}' = \frac{1}{2} \log(1 + \snr_y) - \frac{1}{2} \log(1 + \snr_z').
\end{align}
Any ``good'' codebook sequence for the Gaussian point-to-point channel to $\rvec{y}$, meaning $\rate = C$, attains maximum equivocation for any $\snr_z'$, and is thus also a $(C, d_{\max}')$ codebook sequence for the Gaussian wiretap channel $(\snr_y, \snr_z')$.
\end{lem}

It is well known that $(C, d_{\max}')$ is an achievable rate-equivocation pair for the Gaussian wiretap channel \cite{WynerWireTap,WireTapGaussian}, meaning that full point-to-point rate (capacity rate) does not need to be compromised for the sake of maximum confidentiality. 
This pair is attainable by keeping completely secure only part of the message, as shown in \cite[Ex. a, pp. 408]{CsiszarKorner}.
The above observation applies to \emph{any} ``good'' codebook sequence. It is, on the one hand, a straightforward result, but, on the other hand, very vague, since the secured information is not defined. Note that we are only considering the codebook sequence without having any specific mapping over it. This observation is motivated by the equivalence between the rate-equivocation region and the secret-private region (in which one of the messages is completely secure and the additional rate is for a private message) which has been shown in \cite[Ex. c, pp. 413]{CsiszarKorner}. In our setting we consider \emph{any} ``good'' point-to-point codebook sequence (with a one-to-one mapping), and do not claim only the \emph{existence} of a ``good'' point-to-point sequence that is also $(C, d_{\max}')$. However, we do not show the \emph{existence} of a specific mapping to conclude the definition of the code sequence $(C, d_{\max}')$. Such a mapping is equivalent to finding a mapping from every original message to a product of a completely secure message and a private message. We consider this in the next result.

\begin{thm} \label{thm:equivalence}
For any ``good'' sequence for the Gaussian point-to-point channel to $\rvec{y}$, meaning $\rate = C$, assume a partition of the codewords into $2^{\dim d_{\max}'}$ bins (of approximately equal size), denoted by $W_s$. The next two claims are equivalent:
\begin{enumerate}

\item $W_s$ is completely secure in the equivocation sense (Definition \ref{dfn:perfectSecrecy}).

\item Each bin, with messages denoted by $W_p$, is a ``good'' sequence for a receiver at $\snr_z'$.
\end{enumerate}
\end{thm}

According to the above result, given a ``good'' sequence, which is also a $(C, d_{\max}')$ code for the Gaussian wiretap channel (Lemma \ref{lem:P2Poptimal2WiretapOptimal}), finding a mapping from $W_y$ to $(W_s, W_p)$ (independent) such that $W_s$ is completely secure is equivalent to splitting the codewords into sub-codes that are optimal for the eavesdropper. This approach is exactly the one in Wyner's original work \cite{WynerWireTap}, and also emphasized by Massey in \cite{MasseyWynerWireTap}, wherein the achievability proof the construction of the code sequence is such that the bins of each secure message are ``good'' code sequences to the eavesdropper (saturating the eavesdropper). The above claim extends this observation by claiming that \emph{any} mapping of messages to codewords (alternatively, \emph{any} binning of the codewords) that attains complete secrecy must saturate the eavesdropper, thus supporting the known achievability scheme of Wyner. The open question posed by the above is the existence of such a partition for any ``good'' point-to-point code sequence. Seemingly, that would be shown in the direct part of the proof of the equivalence between the rate-equivocation region and the secure-private region shown in \cite[Ex. c, pp. 413]{CsiszarKorner}; however the proof does not take \emph{any} code sequence of a given rate-equivocation and shows that there exists a mapping to the corresponding secure-private pair code sequence. Alternatively, the proof relies on the achivability proof of the rate-equivocation region \cite[Ex. a, pp. 408]{CsiszarKorner} which actually proves the achivability of a secure-private pair. 
This proof suffices as the goal is to show equivalence between the regions, which is weaker than claiming the existence of a mapping from one code sequence to the other. Moreover, note that the achivability proof relies on the Coding Stuffing Lemma \cite[Lemma 3.18, pp. 334]{CsiszarKorner} which claims that for any set of codewords there exists a subset that is a disjoint union of sets. This disjoint union of sets is the partition of the codebook into bins denoting the secure message. However, the lemma claims the existence of such a subset and not that any subset is comprised of a disjoint union of subsets (and of course is limited to discrete memoryless channels (DMCs) and relies on properties of the image set size \cite[Lemma 3.2, pp. 305]{CsiszarKorner}). To conclude this discussion, the question regarding the existence of such a partition is still an open one, and what we claim is only that given any partition that defines a completely secure message it must saturate the eavesdropper as done in Wyner's achievability proof \cite{WynerWireTap}.

The above result can be extended. Looking at the proof of Theorem \ref{thm:equivalence} one can observe that the fact that the code sequence achieves capacity has no significance in the proof of the equivalence. The important properties are that the code sequence attains maximum equivocation $d_{\max}$ and that we have either complete secrecy for the message $W_s$ or a ``good'' code sequence in each bin. Let us assume a secure-private message pair $(W_s, W_p)$ over an $(\rate(\alpha), d_{\max})$ code sequence, where
\begin{align}
\rate(\alpha) = d_{\max} + \frac{1}{2} \log( 1 + \alpha \snr_z)
\end{align}
for some $\alpha \in [0,1]$. For such a code sequence we can state the following simple corollary (which relies on Corollary \ref{cor:d_maxObservations}, Theorem \ref{thm:specificRates} and Theorem \ref{thm:equivalence}).
\begin{cor} \label{cor:MMSE_Ws}
An $(\rate(\alpha), d_{\max})$ code sequence transmitting the message pair $(W_s,W_p)$ (independent) has the following behavior:
\begin{align}
\MMSE( \rvec{x} ; \gamma |  W_s) = \left\{ \begin{array}{ll} \frac{1}{1 + \gamma}, & \gamma \in [0, \snr_z) \\ 0,& \gamma \geq \snr_z \end{array} \right. .
\end{align}
Thus, the equivalence of Theorem \ref{thm:equivalence} holds also here - $W_s$ is completely secure if and only if the codewords in each bin construct a ``good'' point-to-point code sequence to the eavesdropper (note that for $\alpha < 1$ the mapping of $W_p$ is not a one-to-one mapping over the ``good'' point-to-point code sequences).
\end{cor}
To conclude, we see that any $d_{\max}$ code sequence is built on a ``good'' point-to-point codebook sequence, and that a partition of this codebook sequence to bins defines a completely secure message if and only if the bins themselves are codebook sequences that achieve capacity to the eavesdropper. This is regardless of the rate of the code sequence.


\subsection{Proofs for the Gaussian Wiretap Channel} \label{ssec:proofs}
\begin{IEEEproof}[Proof of Theorem \ref{thm:optimallySecureCode}]
The single-letter expression for the rate-equivocation region of a \emph{degraded} channel (\ref{eq:RateEquivocation}) was given in \cite{WynerWireTap} and extended to the more general case in \cite{CsiszarKorner}. Since we are examining the properties of length-$\dim$ codebooks of the scalar Gaussian wiretap channel, which is always \emph{degraded}, we use the limiting version of (\ref{eq:RateEquivocation}) (see the converse proof in \cite{WynerWireTap} for details).

Using the chain rule of mutual information we have
\begin{align}
\Igen{\rvec{x}_{\dim}, W_y}{ \rvec{y}_{\dim}} & = \Igen{W_y}{\rvec{y}_{\dim}} + \Icond{\rvec{x}_{\dim}}{\rvec{y}_{\dim}}{W_y} \nonumber \\
& = \Igen{\rvec{x}_{\dim}}{\rvec{y}_{\dim}} + \Icond{W_y}{\rvec{y}_{\dim}}{\rvec{x}_{\dim}} \nonumber \\
& = \Igen{\rvec{x}_{\dim}}{\rvec{y}_{\dim}}
\end{align}
where the last transition is due to the Markov chain relationship $W_y - \rvec{x}_{\dim} - \rvec{y}_{\dim}$. Thus, we have
\begin{align}
\Icond{\rvec{x}_{\dim}}{\rvec{y}_{\dim}}{W_y} & = \Igen{\rvec{x}_{\dim}}{\rvec{y}_{\dim}} - \Igen{W_y}{\rvec{y}_{\dim}}.
\end{align}
Similarly and due to the Markov chain condition $W_y - \rvec{x}_{\dim} - \rvec{z}_{\dim}$ we also have
\begin{align}
\Icond{\rvec{x}_{\dim}}{\rvec{z}_{\dim}}{W_y} & = \Igen{\rvec{x}_{\dim}}{\rvec{z}_{\dim}} - \Igen{W_y}{\rvec{z}_{\dim}}.
\end{align}
Subtracting these two equations we obtain the following:
\begin{align} \label{eq:sabtract}
\Icond{\rvec{x}_{\dim}}{\rvec{y}_{\dim}}{W_y} - \Icond{\rvec{x}_{\dim}}{\rvec{z}_{\dim}}{W_y} = \Igen{\rvec{x}_{\dim}}{\rvec{y}_{\dim}} - \Igen{\rvec{x}_{\dim}}{\rvec{z}_{\dim}} - \Igen{W_y}{\rvec{y}_{\dim}} + \Igen{W_y}{\rvec{z}_{\dim}}.
\end{align}
Normalizing by $\dim$ and taking $\dim \to \infty$ we have due to the assumption of an $(\rate, d_{opt})$ codebook (\ref{eq:wiretapDefinition})
\begin{IEEEeqnarray}{rCl}
\IEEEeqnarraymulticol{3}{l}{\Icond{\rvec{x}}{\rvec{y}}{W_y}  - \Icond{\rvec{x}}{\rvec{z}}{W_y}} \nonumber \\
& = & \Igen{\rvec{x}}{\rvec{y}} - \Igen{\rvec{x}}{\rvec{z}} - \rate + \rate - d_{opt} \nonumber \\
& = & \Igen{\rvec{x}}{\rvec{y}} - \Igen{\rvec{x}}{\rvec{z}} - d_{opt} = 0. \label{eq:relatingMutualInformation2Conditioned}
\end{IEEEeqnarray}
Using the I-MMSE relationship \cite{IMMSE} we can conclude that for any $(\rate, d_{opt})$ codebook for the Gaussian wiretap channel
\begin{align} \label{eq:ConclusionDecodingGivenMessgae}
\int_{\snr_z}^{\snr_y} \MMSE( \rvec{x}; \gamma | W_y ) \d \gamma & = 0 \nonumber \\
\MMSE( \rvec{x}; \gamma | W_y ) & = 0, \quad \forall \gamma \geq \snr_z.
\end{align}
This proves (\ref{thm:eq:MMSE_Wbehavior}). It remains to prove (\ref{thm:eq:MMSEbehavior}). This is a simple consequence of the above since we assume that reliable decoding of $W_y$ is possible at SNR at or above $\snr_y$; thus $\forall \gamma \geq \snr_y $
\begin{align} \label{eq:ConclusionFullMessageDecodingAtBob}
\MMSE( \rvec{x}; \gamma ) & = \MMSE( \rvec{x}; \gamma | W_y) \nonumber \\
\MMSE( \rvec{x}; \gamma ) & = 0.
\end{align}
This concludes the proof of the direct part.

For the converse part we now assume that (\ref{thm:eq:MMSEbehavior}) and (\ref{thm:eq:MMSE_Wbehavior}) both hold for a codebook where messages $W_y \in \{1,2,\ldots, 2^{\dim \rate} \}$ are mapped (not necessarily deterministically) to codewords $\rvec{x}_{\dim}$.
From (\ref{thm:eq:MMSEbehavior}) we can conclude reliable decoding of the messages at $\snr_y$ since the MMSE when estimating the codeword $\rvec{x}_{\dim}$ is zero implies that the message can be reliably decoded. Thus, we can use (\ref{eq:wiretapDefinition}) to define $d$ ($d \leq \rate$). The Markov chain relationship $W_y - \rvec{x}_{\dim} - \rvec{y}_{\dim} - \rvec{z}_{\dim}$ still holds and we begin from (\ref{eq:sabtract}). Using (\ref{eq:wiretapDefinition}) we have
\begin{align} \label{eq:sabtract2}
\Icond{\rvec{x}}{\rvec{y}}{W_y}  - \Icond{\rvec{x}}{\rvec{z}}{W_y} & = \Igen{\rvec{x}}{\rvec{y}} - \Igen{\rvec{x}}{\rvec{z}} - d
\end{align}
and we need to show that (\ref{thm:eq:MMSEbehavior}) and (\ref{thm:eq:MMSE_Wbehavior}) guarantee $d = d_{opt}$.

Using the I-MMSE relationship on the left-hand-side of the above equation we have that
\begin{align}
\frac{1}{2}\int_{\snr_z}^{\snr_y} \MMSE( \rvec{x}; \gamma | W_y ) \d \gamma = \Igen{\rvec{x}}{\rvec{y}} - \Igen{\rvec{x}}{\rvec{z}} - d.
\end{align}
From (\ref{thm:eq:MMSE_Wbehavior}) we can conclude that the left-hand-side is zero and
\begin{align} \label{eq:proofTheoremWiretap_reverseConc}
d = \Igen{\rvec{x}}{\rvec{y}} - \Igen{\rvec{x}}{\rvec{z}} = d_{opt}.
\end{align}
This concludes our proof.
\end{IEEEproof}

\begin{IEEEproof}[Proof of Corollary \ref{cor:d_maxObservations}]
In order to prove this corollary we need only to show that the maximum level of equivocation, $d_{\max}$, is a specific case of $d_{opt}$ which also determines the exact behavior of $\MMSE( \rvec{x} | \sqrt{\gamma} \rvec{x} + \rvec{n})$.

Note that
\begin{align} \label{eq:equality_d}
d & \leq \lim_{\dim \to \infty} \frac{1}{\dim} \left[ \Igen{\rvec{x}_{\dim}}{\rvec{y}_{\dim}} - \Igen{\rvec{x}_{\dim}}{\rvec{z}_{\dim}} \right] \nonumber \\
d & \leq \frac{1}{2} \int_{\snr_z}^{\snr_y} \MMSE( \rvec{x}; \gamma ) \d \gamma \leq \frac{1}{2} \log \left( 1 + \snr_y \right) - \frac{1}{2} \log \left( 1 + \snr_z \right) =  d_{\max}
\end{align}
where the last inequality is due to the fact that the MMSE function is upper bounded by the linear MMSE for any SNR. This leads to the following conclusion: if $d = d_{\max}$ then we must have equality in the above equation, meaning $d = d_{opt}$. Thus, the properties of Theorem \ref{thm:optimallySecureCode} define these codes ``if and only if''. Moreover, since
\begin{align}
\frac{1}{2} \int_{\snr_z}^{\snr_y} \MMSE( \rvec{x}; \gamma ) \d \gamma = \frac{1}{2} \log \left( 1 + \snr_y \right) - \frac{1}{2} \log \left( 1 + \snr_z \right)
\end{align}
we can conclude that
\begin{align} \label{eq:ConclusionGaussianCodebook}
\MMSE( \rvec{x}; \gamma ) = \frac{1}{1 + \gamma}, \quad \forall \gamma \in [0, \snr_y)
\end{align}
concluding the direct part of the Corollary. As for the reverse direction, recall that the properties in Theorem \ref{thm:optimallySecureCode} lead to (\ref{eq:proofTheoremWiretap_reverseConc}), that is
\begin{align}
d = \Igen{\rvec{x}}{\rvec{y}} - \Igen{\rvec{x}}{\rvec{z}} = \frac{1}{2} \int_{\snr_z}^{\snr_y} \MMSE( \rvec{x}; \gamma ) \d \gamma;
\end{align}
adding the additional property in this corollary leads to
\begin{align}
d = \frac{1}{2} \log \left( 1 + \snr_y \right) - \frac{1}{2} \log \left( 1 + \snr_z \right) = d_{\max}.
\end{align}
This concludes our proof.
\end{IEEEproof}

\begin{IEEEproof}[Proof of Theorem \ref{thm:specificRates}]
Note that due to the Markov chain condition $W_y - \rvec{x}_{\dim} - \sqrt{\snr} \rvec{x}_{\dim} + \rvec{n}_{\dim}$ we have that for any $\snr$
\begin{align}
\Igen{W_y}{\sqrt{\snr} \rvec{x} + \rvec{n}} = \Igen{\rvec{x}}{\sqrt{\snr} \rvec{x} + \rvec{n}} - \Icond{\rvec{x}}{\sqrt{\snr} \rvec{x} + \rvec{n} }{W_y}
\end{align}
and using the I-MMSE relationship
\begin{align} \label{eq:rateOfWiretap}
\Igen{W_y}{\sqrt{\snr} \rvec{x} + \rvec{n}} =  \frac{1}{2} \int_0^{\snr} \left[ \MMSE\left( \rvec{x}; \gamma \right) - 
\MMSE \left( \rvec{x} ; \gamma | W_y \right) \right] \d \gamma.
\end{align}
Notice that for any $\gamma$
\begin{align}
\MMSE\left( \rvec{x}; \gamma  \right) \geq \MMSE \left( \rvec{x} ; \gamma |  W_y \right).
\end{align}
This is due to the concavity of the MMSE function in the input distribution \cite[Corollary 1]{FunctionalPropertiesMMSE} (note that the result extends to the limit when $\dim \to \infty$ simply by taking the limit in \cite[Equation (10)]{FunctionalPropertiesMMSE}). The distribution of $\rvec{x}$ can be written as $\sum_i P_{w_i} P_{\rvec{x} | w_i}$ and the right-hand-side is the expectation over
\begin{align}
\MMSE \left( \rvec{x}_{w_i} ; \gamma |  W_y = w_i \right) = \MMSE( \rvec{x}_{w_i} | \sqrt{\gamma} \rvec{x}_{w_i} + \rvec{n}, W_y = w_i).
\end{align}
As such the integrand in (\ref{eq:rateOfWiretap}) is non-negative for all $\gamma$.

If we require complete secrecy, we have that
\begin{align}
\Igen{W_y}{\sqrt{\snr_z} \rvec{x} + \rvec{n}} & = 0 \nonumber \\
\frac{1}{2} \int_0^{\snr_z} \left[ \MMSE\left( \rvec{x}; \gamma \right) - 
\MMSE \left( \rvec{x}; \gamma | W_y \right) \right] \d \gamma & = 0.
\end{align}
Thus, we can conclude that
\begin{align}
\MMSE\left( \rvec{x}; \gamma \right)  = \MMSE \left( \rvec{x}; \gamma |  W_y \right), \quad \forall \gamma \in [0, \snr_z).
\end{align}
From the reliable decoding at $\rvec{y}$ and the above equality we can conclude that (\ref{thm:eq:PerfectSecrecyReliableComm}) holds.\\
In the reverse direction, setting the above equality leads exactly to reliable communication at rate $\rate$ with complete secrecy.

If we require $\rate_{\max}$, then
\begin{align}
\Igen{W_y}{\sqrt{\snr_y} \rvec{x} + \rvec{n}} & = \Igen{\rvec{x}}{\rvec{y}} \nonumber \\
\frac{1}{2} \int_0^{\snr_y} \left[ \MMSE\left( \rvec{x}; \gamma  \right) -
\MMSE \left( \rvec{x} ; \gamma |  W_y \right) \right] \d \gamma & = \frac{1}{2} \int_0^{\snr_y} \MMSE( \rvec{x}; \gamma ) \d \gamma
\end{align}
from which we can conclude that
\begin{align}
\MMSE \left( \rvec{x}; \gamma | W_y \right) = 0, \quad \forall \gamma \geq 0
\end{align}
as expected in a one-to-one mapping of message to codeword. This again holds in the reverse direction. \\
Finally, if we further specify that $d = d_{\max}$ the exact behavior is determined from Corollary \ref{cor:d_maxObservations}. This concludes our proof.
\end{IEEEproof}

\begin{IEEEproof}[Proof of Lemma \ref{lem:P2Poptimal2WiretapOptimal}]
We first prove that any ``good'' point-to-point sequence is a $(C, d_{\max}')$ sequence for the Gaussian wiretap channel. This is simply observed from (\ref{eq:sabtract2}) where the left-hand-side is zero due to the one-to-one mapping of the message to a codeword, and we remain with
\begin{align}
d = \Igen{\rvec{x}}{\rvec{y}} - \Igen{\rvec{x}}{\rvec{z}'}
\end{align}
which equals $d_{\max}'$ since this is a ``good'' sequence and thus (\ref{cor:eq:MMSEbehavior}) holds \cite{EXIT,StatisticalPhysics}. This concludes the proof.
\end{IEEEproof}

\begin{IEEEproof}[Proof of Theorem \ref{thm:equivalence}]
Assume that the first claim holds and we have a partition to $2^{\dim d_{\max}'}$ bins that is completely secure, meaning
$\Igen{ W_s}{\rvec{z}'} = 0$.
Thus
\begin{align} \label{eq:proof:wiretapEquivalenceClaim}
\Icond{\rvec{x}}{\rvec{z}'}{W_s} & = \Igen{\rvec{x}}{\rvec{z}'} - \Igen{W_s}{\rvec{z}'} \nonumber \\
& = \frac{1}{2} \log(1 + \snr_z').
\end{align}
Due to the following upper bound
\begin{align}
\Icond{\rvec{x}}{\rvec{z}'}{W_s = w} \leq \frac{1}{2} \log(1 + \snr_z')
\end{align}
we can conclude that for any $W_s = w$, meaning, for any bin
\begin{align}
\Icond{\rvec{x}}{\rvec{z}'}{W_s = w} = \frac{1}{2} \log(1 + \snr_z').
\end{align}
Finally, using the result of Corollary \ref{cor:d_maxObservations} we have that
\begin{align}
\MMSE( \rvec{x}; \gamma | W_s) = 0, \quad  \forall \gamma \geq \snr_z'
\end{align}
implying that given the bin, reliable decoding is possible at $\rvec{z}'$. This concludes the first direction.
The other direction is even simpler since a ``good'' sequence in every bin means
\begin{align} \label{eq:proof:equivalence1}
\Icond{\rvec{x}}{\rvec{z}'}{W_s = w} & = \frac{1}{2} \log(1 + \snr_z').
\end{align}
Thus, we can immediately conclude that $\Igen{W_s}{\rvec{z}'} = 0$ via (\ref{eq:proof:wiretapEquivalenceClaim}). This concludes the proof.
\end{IEEEproof}

\section{The Gaussian Broadcast Channel} \label{sec:GaussianBC}

\subsection{Model and Definitions} \label{ssec:BC:ModelAndDef}
Assume a scalar Gaussian BC, over which we transmit length-$\dim$ codewords
\begin{align} \label{eq:modelGaussianBC}
\rvec{y}_{\dim} = \sqrt{ \snr_y } \rvec{x}_{\dim} + \rvec{n_1}_{\dim} \nonumber \\
\rvec{z}_{\dim} = \sqrt{ \snr_z } \rvec{x}_{\dim} + \rvec{n_2}_{\dim}
\end{align}
where $\rvec{x}_{\dim}$ is the length-$\dim$ transmitted codeword. $\rvec{n_1}_{\dim}$ and $\rvec{n_2}_{\dim}$ are standard additive Gaussian noise vectors. Without loss of generality we assume that $\snr_y > \snr_z$, making $\rvec{y}$ the stronger receiver, which can decode both messages.
As the capacity region of the above channel depends only on the marginals of the joint conditional distribution of the outputs given the input \cite{CoverThomas}, we may assume, without loss of generality, $\rvec{n_1}_{\dim} = \rvec{n_2}_{\dim} = \rvec{n}_{\dim}$.


An $(\rate_y, \rate_z)$ code sequence, $( \C, f )$, for this channel must reliably transmit the message, $W_y$, of cardinality $2^{\dim \rate_y}$, to $\rvec{y}$ and reliably transmit the message, $W_z$, of cardinality $2^{\dim \rate_z}$, to $\rvec{z}$.
That is, there exists a sequence of encoding functions such that $\rvec{x}_{\dim} = f_{\dim}( W_y, W_z)$, where $W_y \in \{ 1, 2, \ldots, 2^{\dim \rate_y} \}$ and $W_z \in \{ 1, 2, \ldots, 2^{\dim \rate_z} \}$ such that the error probability when decoding $W_z$ from $\rvec{z}_{\dim}$ and the error probability when decoding $W_y$ from $\rvec{y}_{\dim}$, go to zero as $\dim \to \infty$.
These requirements can be written as follows:
\begin{align} \label{eq:wiretapDefinition}
\lim_{\dim \to \infty} \frac{1}{\dim} \Igen{W_y}{\rvec{y}_{\dim}} & = \rate_y \nonumber \\
\lim_{\dim \to \infty} \frac{1}{\dim} \Igen{W_z}{\rvec{z}_{\dim}} & = \rate_z.
\end{align}

The capacity region of the channel is well known \cite{CoverThomas}:
\begin{align}
C = \left\{ \begin{array}{l} (\rate_y, \rate_z): \\
0 \leq \rate_y \leq \Icond{\resc{x}}{\resc{y}}{\resc{u}} \\
0 \leq \rate_z \leq \Igen{\resc{u}}{\resc{z}}
\end{array}
\right. .
\end{align}
where $\resc{u}$ is an auxiliary random variable such that $\resc{u} - \resc{x} - \resc{y} - \resc{z}$ forms a Markov chain. In the Gaussian setting considered here the capacity region is well-known \cite{Bergmans_Converse}, and can is obtained by choosing $\resc{x} = \resc{u}+\resc{v}$ where $\resc{v}$ and $\resc{u}$ are independent Gaussian random variables with variances that sum to one (Bergmans \cite{Bergmans_Converse} has provided a converse proof that does not rely on the single-letter expression \cite{Gallager_Converse} using the entropy power inequality (EPI). A simpler proof based on the single-letter expression using the EPI was shown in \cite{ElGamalLectureNotes}, and a proof using the I-MMSE relationship was given in \cite{PROP_full}).

Given these requirements and due to the data-processing lemma we have the following inequality
\begin{align}
\lim_{\dim \to \infty} \frac{1}{\dim}\Igen{ \rvec{x}_{\dim}}{\rvec{y}_{\dim}} \geq  \rate_y + \rate_z = \Igen{W_y,W_z}{\rvec{y}_{\dim}}.
\end{align}
Equality holds when we have a deterministic encoder.

\subsection{Main Results} \label{ssec:BCmainResults}

We first wish to understand what are the implications of reliable decoding in terms of the MMSE behavior. We consider any code sequence designed for reliable communication of the pair $(W_y, W_z)$ to the two receivers. The sequences can be either ``good'', capacity achieving, sequences or ``bad'' sequences, that do not necessarily achieve capacity.

\begin{thm} \label{thm:badCodeGaussianBC}
Consider a code sequence, transmitting a message pair $(W_y, W_z)$, at rates $(\rate_y, \rate_z)$, over the Gaussian BC. $W_z$ can be reliably decoded
from $\rvec{z}$ if and only if 
\begin{align} \label{eq:thm:ReliableDecoding}
\MMSE( \rvec{x} ; \gamma | W_z) = \MMSE( \rvec{x} ; \gamma ), \quad \forall \gamma \geq \snr_z.
\end{align}
\end{thm}

The above theorem formally states a very obvious observation which is that once $W_z$ can be decoded, it provides no additional information to the estimation of the transmitted codeword, beyond the information in the output. We strengthen this insight by proving that this property is sufficient for reliable decoding.

In \cite[Theorem 1]{FunctionalPropertiesMMSE} it was shown that the MMSE function is a concave function in the input distribution. Moreover, for the additive Gaussian channel is was shown \cite[Theorem 2]{FunctionalPropertiesMMSE} that this is actually a strict concavity. However, these results were shown for finite dimension, and although the concavity result transfers quite simply to the limit, the condition for strict concavity does not. As such the above does not contradict these results.

Our main result is an extension of the result given in \cite{StatisticalPhysics}, where it was shown that a typical code from the hierarchical code ensemble (which achieves capacity) designed for a given Gaussian BC has a specific MMSE behavior. We extend this result to any ``good'' code sequence as follows:
\begin{thm} \label{thm:GaussianBC_codebook}
Any ``good'' code sequence for the Gaussian BC, with a rate pair that can be depicted as follows:
\begin{align} \label{eq:thm:BC:capacityPoint}
(\rate_y, \rate_z) = \left( \frac{1}{2} \log \left( 1 + \beta \snr_y \right), \frac{1}{2} \log \left( \frac{1 + \snr_z}{1 + \beta \snr_z} \right) \right)
\end{align}
for some $\beta \in [0,1]$, has the MMSE behavior of a Gaussian superposition codebook as $\dim \to \infty$, that is
\begin{align} \label{eq:thm:BC:MMSEbehavior}
\MMSE( \rvec{x}; \snr ) = \left\{ \begin{array}{ll} \frac{1}{1 + \snr}, & \snr \in [0, \snr_z) \\ \frac{\beta}{1 + \beta \snr}, & \snr \in [\snr_z, \snr_y) \\ 0, & \snr \geq \snr_y \end{array} \right. .
\end{align}
\end{thm}

Note that the above behavior holds for any ``good'' code sequence for the Gaussian BC. This includes also codes designed for decoding schemes such as ``dirty paper coding'', in which case the decoding at $\rvec{y}$ does not require the reliable decoding of the known ``interference'' (the part of the codeword that carries the information of $W_z$), but simply encodes the desired messages against that ``interference''. As such, one does not expect such a scheme to have the same MMSE behavior as a superposition code scheme, where the decoding is in layers: first the ``interference'' and only after its removal, the reliable decoding of the desired message.

The above theorem considers only the codebook sequence, and not the mapping of messages to codewords. Moreover, from the proof one can see that (as known) a deterministic mapping is required for any ``good'' code sequence.

\begin{thm} \label{thm:MMSEconditionedPropertiesBC}
Any code sequence complying with (\ref{eq:thm:ReliableDecoding}) and (\ref{eq:thm:BC:MMSEbehavior}) is a reliable code sequence for the Gaussian BC depicted in (\ref{eq:modelGaussianBC}) and will have the following property:
\begin{align} \label{eq:thm:MMSEconditionedWz_inequality}
\MMSE( \rvec{x}; \gamma | W_z) \geq \frac{\beta}{1 + \beta \gamma}, \quad \forall \gamma \in [0, \snr_z].
\end{align}
Moreover, such a code sequence is ``good'' if and only if on top of (\ref{eq:thm:ReliableDecoding}) and (\ref{eq:thm:BC:MMSEbehavior}) it has a deterministic mapping from $(W_y,W_z)$ to the transmitted codeword and
\begin{align} \label{eq:thm:MMSEconditionedWz_equality}
\MMSE( \rvec{x}; \gamma | W_z) = \frac{\beta}{1 + \beta \gamma}, \quad \forall \gamma \in [0, \snr_z].
\end{align}
\emph{I.e.}, (\ref{eq:thm:MMSEconditionedWz_inequality}) is satisfied with equality.
\end{thm}

\begin{figure}[h]
\begin{center}
\setlength{\unitlength}{.1cm}
    \includegraphics[width=1.1\textwidth]{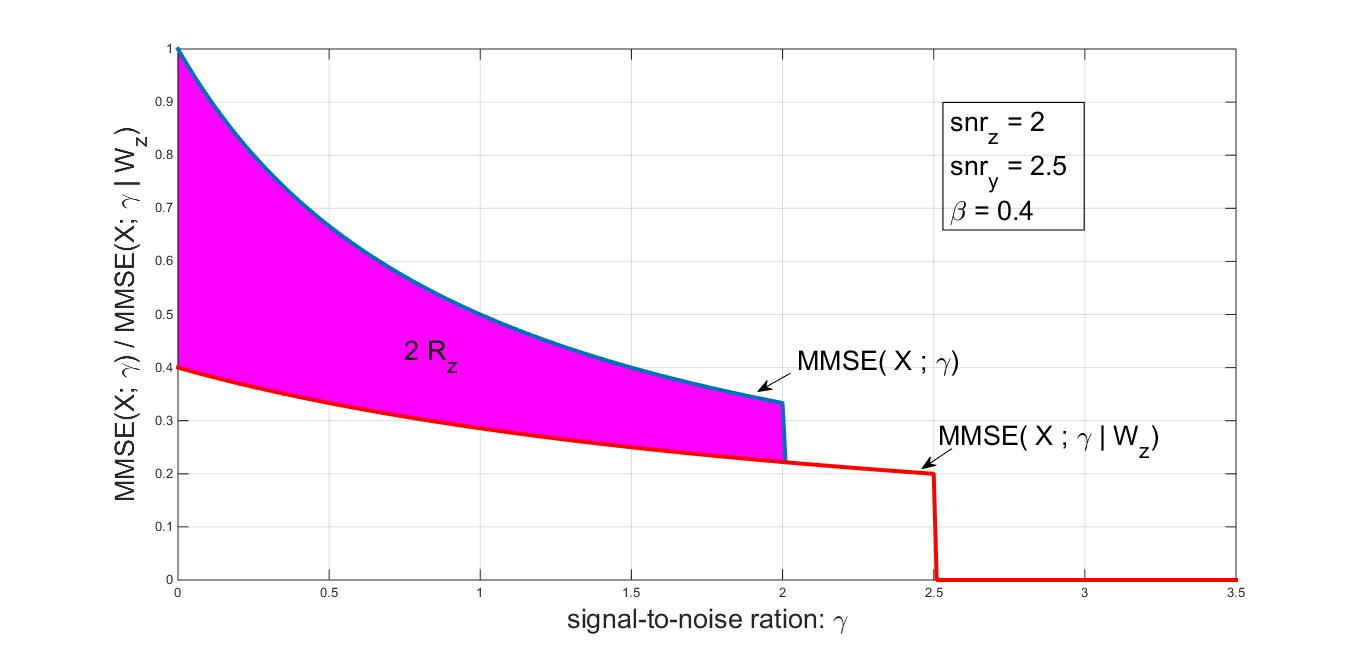}
    \caption{In the above figure we consider the behavior of $\MMSE(\rvec{x}; \gamma)$ (in blue) and $\MMSE( \rvec{x} ;\gamma | W_z)$ (in red) required from a ``good'' code sequence for the Gaussian BC. Twice $\rate_z$ is marked as the area between these two function (in magenta).}
 \label{figure:BC_opt}
\end{center}
\end{figure}

Figure \ref{figure:BC_opt} depicts the result of Theorem \ref{thm:MMSEconditionedPropertiesBC} for ``good'' code sequences.

\subsection{Proofs for the Gaussian BC} \label{ssec:ProofsBC}
\begin{IEEEproof}[Proof of Theorem \ref{thm:badCodeGaussianBC}]
First notice that due to the data-processing lemma we have that
\begin{align}
\Igen{W_z}{\sqrt{\snr_z} \rvec{x}_{\dim} + \rvec{n}_{\dim}} \leq \Igen{W_z}{\sqrt{\snr} \rvec{x}_{\dim} + \rvec{n}_{\dim}} \nonumber
\end{align}
for any $\snr \geq \snr_z$. Since we have reliable decoding of $W_z$ at $\rvec{z}$ and using Fano's inequality we have
that as $\dim \to \infty$
\begin{align}
\Igen{W_z}{\sqrt{\snr_z} \rvec{x} + \rvec{n}} = \rate_z
\end{align}
and
\begin{align} \label{proof:thm:badCodeBC:equalityMI}
\Igen{W_z}{\sqrt{\snr} \rvec{x} + \rvec{n}} = \Igen{W_z}{\sqrt{\snr_z} \rvec{x} + \rvec{n}}
\end{align}
for all $\snr \geq \snr_z$. \\
Now, consider the chain rule of mutual information,
\begin{IEEEeqnarray}{rCl}
\IEEEeqnarraymulticol{3}{l}{\Igen{ \rvec{x}_{\dim}, W_z}{\sqrt{\snr} \rvec{x}_{\dim} + \rvec{n}_{\dim}}} \nonumber \\
& = & \Igen{W_z}{\sqrt{\snr} \rvec{x}_{\dim} + \rvec{n}_{\dim}} + \Icond{\rvec{x}_{\dim}}{\sqrt{\snr} \rvec{x}_{\dim} + \rvec{n}_{\dim}}{W_z} \nonumber \\
& = & \Igen{\rvec{x}_{\dim}}{\sqrt{\snr} \rvec{x}_{\dim} + \rvec{n}_{\dim}} + \Icond{W_z}{\sqrt{\snr} \rvec{x}_{\dim} + \rvec{n}_{\dim}}{\rvec{x}_{\dim}} \nonumber \\
& = & \Igen{\rvec{x}_{\dim}}{\sqrt{\snr} \rvec{x}_{\dim} + \rvec{n}_{\dim}}
\end{IEEEeqnarray}
where the first and second transitions both result from applying the chain rule of mutual information on $\Igen{ \rvec{x}_{\dim}, W_z}{\sqrt{\snr} \rvec{x}_{\dim} + \rvec{n}_{\dim}}$.
The last equality is due to the Markov chain relationship $(W_z, W_y) - \rvec{x}_{\dim} - \sqrt{\snr} \rvec{x}_{\dim} + \rvec{n}_{\dim}$. Thus, we have
\begin{align} \label{eq:chainMI}
\Igen{W_z}{\sqrt{\snr} \rvec{x}_{\dim} + \rvec{n}_{\dim}} = \Igen{\rvec{x}_{\dim}}{\sqrt{\snr} \rvec{x}_{\dim} + \rvec{n}_{\dim}} -
\Icond{\rvec{x}_{\dim}}{\sqrt{\snr} \rvec{x}_{\dim} + \rvec{n}_{\dim}}{W_z}.
\end{align}
The above holds for all $\dim$.
Taking into account the equality in (\ref{proof:thm:badCodeBC:equalityMI}), which holds when $\dim \to \infty$, we have that for any $\snr \geq \snr_z$
\begin{align}
\Igen{\rvec{x}}{\sqrt{\snr} \rvec{x} + \rvec{n}} - \Igen{\rvec{x}}{\sqrt{\snr_z} \rvec{x} + \rvec{n}} = \Icond{\rvec{x}}{\sqrt{\snr} \rvec{x} + \rvec{n}}{W_z} -  \Icond{\rvec{x}}{\sqrt{\snr_z} \rvec{x} + \rvec{n}}{W_z}
\end{align}
which in terms of the I-MMSE relationship \cite{IMMSE} gives us the following:
\begin{align}
\int_{\snr_z}^{\snr} \MMSE( \rvec{x} ; \gamma) \d \gamma = \int_{\snr_z}^{\snr} \MMSE( \rvec{x} ; \gamma | W_z) \d \gamma.
\end{align}
Due to the concavity of the MMSE in the input distribution \cite{FunctionalPropertiesMMSE} (which holds also in the $\lim$ or the $\limsup$) we have that
\begin{align}
\MMSE( \rvec{x} ; \gamma) \geq  \MMSE( \rvec{x} ; \gamma | W_z), \quad \forall \gamma \geq 0
\end{align}
and we can conclude that
\begin{align}
\MMSE( \rvec{x} ; \gamma) =  \MMSE( \rvec{x} ; \gamma | W_z), \quad \forall \gamma \geq \snr_z.
\end{align}
This concludes the direct part of the proof.

For the converse, we assume that (\ref{eq:thm:ReliableDecoding}) holds. Applying the I-MMSE relationship on (\ref{eq:chainMI}) we have that
\begin{align}
\Igen{W_z}{\sqrt{\snr} \rvec{x} + \rvec{n}} = \frac{1}{2} \int_0^{\snr} \MMSE( \rvec{x}; \gamma) - \MMSE( \rvec{x}; \gamma| W_z ) \d \gamma.
\end{align}
Due to (\ref{eq:thm:ReliableDecoding}) we have that for every $\snr \geq \snr_z$
\begin{align}
\Igen{W_z}{\sqrt{\snr} \rvec{x} + \rvec{n}} 
& = \Igen{W_z}{\sqrt{\snr_z} \rvec{x} + \rvec{n}}.
\end{align}
For $\snr \to \infty$ we have that
\begin{align}
\lim_{\snr \to \infty} \Igen{W_z}{\sqrt{\snr} \rvec{x} + \rvec{n}} = \Igen{W_z}{\rvec{x}} = H( W_z) = \rate_z. \nonumber
\end{align}
As such we have that
\begin{align}
\Igen{W_z}{\sqrt{\snr_z} \rvec{x} + \rvec{n}} = H(W_z) = \rate_z
\end{align}
meaning reliable decoding of $W_z$ from $\rvec{z}$. This concludes the proof.
\end{IEEEproof}

\begin{IEEEproof}[Proof of Theorem \ref{thm:GaussianBC_codebook}]
Using (\ref{eq:chainMI}) we have
\begin{align}
\Icond{\rvec{x}_{\dim}}{\rvec{z}_{\dim}}{W_z} & = \Igen{\rvec{x}_{\dim}}{\rvec{z}_{\dim}} - \Igen{W_z}{\rvec{z}_{\dim}} \nonumber \\
& = \Igen{\rvec{x}_{\dim}}{\rvec{z}_{\dim}} - \left[ \Hen{W_z} - \HenC{W_z}{\rvec{z}_{\dim}} \right] \nonumber \\
& = \Igen{\rvec{x}_{\dim}}{\rvec{z}_{\dim}} - \dim \rate_z + \HenC{W_z}{\rvec{z}_{\dim}}.
\end{align}
Using Fano's inequality and the trivial upper bound on the mutual information we have
\begin{align}
\Icond{\rvec{x}_{\dim}}{\rvec{z}_{\dim}}{W_z} & = \Igen{\rvec{x}_{\dim}}{\rvec{z}_{\dim}} - \dim \rate_z + \HenC{W_z}{\rvec{z}_{\dim}} \nonumber \\
& \leq \frac{n}{2} \log \left( 1 + \snr_z \right) + \Hbin{ P^{\dim}(e) } + P^{\dim}(e) \log \left( 2^{\dim \rate_z} - 1 \right) - \dim \rate_z \nonumber
\end{align}
where $\Hbin{\cdot}$ denotes the binary entropy.
Dividing the above by $\dim$ and denoting
\begin{align}
\delta_{\dim} = \frac{1}{\dim} \left[ \Hbin{ P^{\dim}(e) } + P^{\dim}(e) \log \left( 2^{\dim \rate_z} - 1 \right) \right]
\end{align}
we have
\begin{align}
\frac{1}{\dim}\Icond{\rvec{x}_{\dim}}{\rvec{z}_{\dim}}{W_z} & \leq \frac{1}{2} \log \left( 1 + \snr_z \right) - \rate_z + \delta_{\dim} \nonumber
\end{align}
where $\delta_{\dim} \to 0$ as $\dim \to \infty$.
Since we assume (\ref{eq:thm:BC:capacityPoint}), that is,
\begin{align} \label{eq:canBeArbitrary}
\rate_z = \frac{1}{2} \log \left( \frac{1 + \snr_z}{1 + \beta \snr_z} \right),
\end{align}
we have
\begin{align} \label{eq:BC:beforeIMMSE}
\frac{1}{\dim}\Icond{\rvec{x}_{\dim}}{\rvec{z}_{\dim}}{W_z} & \leq  \frac{1}{2} \log \left( 1 + \beta \snr_z \right) + \delta_{\dim}.
\end{align}
Recall the conditional I-MMSE relationship:
\begin{align}
\frac{1}{\dim}\Icond{\rvec{x}_{\dim}}{\rvec{z}_{\dim}}{W_z} & = \frac{1}{2} \int_{0}^{\snr_z} \MMSE( \rvec{x}_{\dim} ; \gamma | W_z) \d \gamma.
\end{align}
Using this in (\ref{eq:BC:beforeIMMSE}) and taking the limit as $\dim \to \infty$ we obtain
\begin{align}
\frac{1}{2} \int_0^{\snr_z} \MMSE( \rvec{x}; \gamma | W_z ) \d \gamma & \leq  \frac{1}{2} \log \left( 1 + \beta \snr_z \right).
\end{align}
Due to the ``single crossing point'' property, given in Theorem \ref{thm:ScalarUniqueCrossingPoint}, we can conclude that the crossing point, if it exists, occurred within $[0, \snr_z)$, and thus, at $\snr_z$ we have the following upper bound:
\begin{align} \label{eq:BC:upperBound}
\MMSE( \rvec{x} ; \snr_z |  W_z ) \leq \frac{\beta}{1 + \beta \snr_z}.
\end{align}
According to Theorem \ref{thm:badCodeGaussianBC}
\begin{align} \label{eq:BC:decodingRequirement}
\MMSE( \rvec{x}; \snr_z |  W_z ) = \MMSE( \rvec{x}; \snr_z )
\end{align}
since $W_z$ can be reliably decoded from $\sqrt{ \snr_z } \rvec{x} + \rvec{n}$.
In \cite[Theorem 3]{IMMSEtradeoff} the following optimization problem has been investigated
\begin{align}
\max & \quad \Igen{ \rvec{x} }{\sqrt{\snr_y}\rvec{x} + \rvec{n} }  \\ \label{eq:BCproofOptimization}
\textrm{s.t.} & \quad \MMSE( \rvec{x}; \snr_z ) \leq \frac{\beta}{1 + \beta \snr_z} .
\end{align}
It was shown that the solution is
\begin{align}
\max \Igen{ \rvec{x} }{\sqrt{\snr_y}\rvec{x} + \rvec{n} } =  \frac{1}{2} \log \left( 1 + \beta \snr_y \right) + \frac{1}{2} \log \left( \frac{1 + \snr_z}{1 + \beta \snr_z} \right).
\end{align}
Recall that we have a ``good'' code sequence for the Gaussian BC, meaning that
\begin{align}
\rate_y + \rate_z = \frac{1}{2} \log \left( 1 + \beta \snr_y \right) + \frac{1}{2} \log \left( \frac{1 + \snr_z}{1 + \beta \snr_z} \right).
\end{align}
Given the above and the data-processing inequality we can conclude that in order to obtain the sum-capacity the mapping must be deterministic and we must obtain the maximum for $\Igen{ \rvec{x} }{\sqrt{\snr_y}\rvec{x} + \rvec{n} }$.
Thus, the capacity point of the Gaussian BC will have the following two properties:
\begin{align}
\Igen{ \rvec{x} }{\sqrt{\snr_y}\rvec{x} + \rvec{n} } & = \frac{1}{2} \log \left( 1 + \beta \snr_y \right) + \frac{1}{2} \log \left( \frac{1 + \snr_z}{1 + \beta \snr_z} \right) \\
\MMSE( \rvec{x}; \snr_z ) & = \frac{\beta}{1 + \beta \snr_z}
\end{align}
and according to \cite[Theorem 6]{IMMSEtradeoff} the MMSE and mutual information behavior of this code is fully defined (for every SNR) and is that of the Gaussian two-layered superposition codebook (optimally designed for the Gaussian BC with a rate splitting coefficient $\beta$). Finally, note that we have shown that the mutual information $\Igen{ \rvec{x} }{\sqrt{\snr_y}\rvec{x} + \rvec{n} }$ cannot exceed the sum-rate. This means that the encoder must be deterministic in order to obtain capacity.
This concludes our proof.
\end{IEEEproof}

\begin{IEEEproof}[Proof of Theorem \ref{thm:MMSEconditionedPropertiesBC}]
From the assumptions we first conclude that the code sequence is a reliable code sequence for the Gaussian BC. This is due to Theorem \ref{thm:badCodeGaussianBC} that ensures the reliable decoding of $W_z$ from $\rvec{z}$ and the fact that $\MMSE( \rvec{x} ; \gamma)$ is zero at $\snr_y$ meaning the entire codeword can be reliably decoded from $\rvec{y}$. Moreover, from Theorem \ref{thm:badCodeGaussianBC} and the ``single crossing point'' property given in Theorem \ref{thm:ScalarUniqueCrossingPoint} we can conclude that, since
\begin{align}
\MMSE( \rvec{x}; \snr_z | W_z) = \frac{\beta}{1 + \beta \snr_z}
\end{align}
we must have that
\begin{align} \label{eq:behaviorOfMMSE_Wz_BC}
\MMSE( \rvec{x}; \gamma | W_z) \geq \frac{\beta}{1 + \beta \gamma}, \quad \forall \gamma \in [0, \snr_z).
\end{align}
It remains to show that the code sequence is ``good'', if and only if, on top of (\ref{eq:thm:ReliableDecoding}) and (\ref{eq:thm:BC:MMSEbehavior}) the code sequence has a deterministic mapping and we have equality in (\ref{eq:behaviorOfMMSE_Wz_BC}) for all $\gamma \in [0, \snr_z)$.

If the code sequence is ``good'' then the rate pair is of the form shown in (\ref{eq:thm:BC:capacityPoint}) and the behavior of $\MMSE(\rvec{x} ; \gamma)$ is as proved in Theorem \ref{thm:GaussianBC_codebook} (and the mapping must be deterministic). According to Theorem \ref{thm:badCodeGaussianBC} we also have (\ref{eq:thm:ReliableDecoding}). Due to the chain rule of mutual information (\ref{eq:chainMI}) we have
\begin{align} \label{eq:obtainingRatez}
\rate_z & = \Igen{W_z}{\rvec{z}} \nonumber \\
& = \Igen{\rvec{x}}{\sqrt{\snr_z} \rvec{x} + \rvec{n} } - \Icond{\rvec{x}}{\sqrt{\snr_z} \rvec{x} + \rvec{n} }{W_z} \nonumber \\
& = \frac{1}{2} \int_0^{\snr_z} \MMSE( \rvec{x}; \gamma) - \MMSE( \rvec{x}; \gamma | W_z) \d \gamma
\end{align}
and thus
\begin{align}
\frac{1}{2} \int_0^{\snr_z} \MMSE( \rvec{x}; \gamma | W_z) \d \gamma = \frac{1}{2} \log (1 + \beta \snr_z).
\end{align}
Due to (\ref{eq:behaviorOfMMSE_Wz_BC}) this holds only if we have equality for all $\gamma \in [0, \snr_z)$.
This proves the direct part.

As claimed above, conditions (\ref{eq:thm:ReliableDecoding}) and (\ref{eq:thm:BC:MMSEbehavior}) suffice for the code sequence to be a reliable code sequence for the Gaussian BC.
Similarly to the approach used to obtain (\ref{eq:obtainingRatez}) we can obtain an expression for $\rate_y$:
\begin{align}
\rate_y & = \frac{1}{2} \int_0^{\snr_y} \MMSE( \rvec{x}; \gamma) - \MMSE( \rvec{x}; \gamma | W_y) \d \gamma.
\end{align}
Due to (\ref{eq:thm:BC:MMSEbehavior}) we have
\begin{align} \label{eq:RatesGaussianBCz}
\rate_z & = \frac{1}{2} \log(1 + \snr_z) - \frac{1}{2} \int_0^{\snr_z} \MMSE( \rvec{x}; \gamma | W_z) \d \gamma \intertext{and}
\rate_y & = \frac{1}{2} \log \left( \frac{(1 + \snr_z)(1 + \beta \snr_y)}{1 + \beta \snr_z }\right) - \frac{1}{2} \int_0^{\snr_y} \MMSE( \rvec{x}; \gamma | W_y) \d \gamma \nonumber \\
& = \frac{1}{2} \log \left( \frac{(1 + \snr_z)(1 + \beta \snr_y)}{1 + \beta \snr_z }\right) - \frac{1}{2} \int_0^{\snr_z} \MMSE( \rvec{x}; \gamma | W_y) \d \gamma \label{eq:RatesGaussianBCy}
\end{align}
where the second equality in (\ref{eq:RatesGaussianBCy}) is due deterministic mapping of messages to codewords. In such a case, since $W_z$ is reliably decoded at $\rvec{z}$, the knowledge of $W_y$ is sufficient to obtain the exact transmitted codeword, meaning the MMSE function will go to zero.


Using the chain rule of mutual information we also have
\begin{align}
\Igen{W_z, W_y}{\rvec{z}} & = \Igen{W_y}{\rvec{z}} + \Icond{W_z}{\rvec{z}}{W_y} \nonumber \\
& = \Igen{W_y}{\rvec{z}, W_z} + \Icond{W_z}{\rvec{z}}{W_y} \nonumber \\
& = \Icond{W_y}{\rvec{z}}{W_z} + \Icond{W_z}{\rvec{z}}{W_y}
\end{align}
where the second transition is due to the reliable decoding of $W_z$ from $\rvec{z}$ and the final transition is due to the independence of the two transmitted messages. Using the assumption of a deterministic mapping of messages to codewords the above reduces to
\begin{align}
\Igen{\rvec{x}}{\rvec{z}} & = \Icond{\rvec{x}}{\rvec{z}}{W_z} + \Icond{\rvec{x}}{\rvec{z}}{W_y}.
\end{align}
Applying the I-MMSE relationship we obtain the following relation:
\begin{align}
\frac{1}{2} \int_0^{\snr_z} \MMSE( \rvec{x}; \gamma) \d \gamma & = \frac{1}{2} \int_0^{\snr_z} \MMSE( \rvec{x}; \gamma | W_z) \d \gamma + \frac{1}{2} \int_0^{\snr_z} \MMSE( \rvec{x}; \gamma | W_y) \d \gamma. \nonumber
\end{align}
Note that this relationship holds only for the integration up to $\snr_z$ due to the reliable decoding of $W_z$ at that point, and does not imply any such equality between the integrands (the MMSE quantities).
Given (\ref{eq:thm:BC:MMSEbehavior}) and (\ref{eq:thm:MMSEconditionedWz_equality}) we have that
\begin{align}
\frac{1}{2} \int_0^{\snr_z} \MMSE( \rvec{x}; \gamma | W_y) \d \gamma & = \frac{1}{2} \int_0^{\snr_z} \MMSE( \rvec{x}; \gamma) \d \gamma  - \frac{1}{2} \int_0^{\snr_z} \MMSE( \rvec{x}; \gamma | W_z) \d \gamma \nonumber \\
& = \frac{1}{2} \log(1 + \snr_z) - \frac{1}{2} \log(1 + \beta \snr_z).
\end{align}
Substituting the above in (\ref{eq:RatesGaussianBCy}) we obtain
\begin{align}
\rate_y & = \frac{1}{2} \log(1 + \beta \snr_y)
\end{align}
thus concluding the proof.

\end{IEEEproof}


\section{Gaussian BC with Confidential Messages} \label{sec:GaussianBCC}
\subsection{Model and Definitions} \label{ssec:BCCmodelDef}
Combining the Gaussian wiretap channel and the Gaussian BC we obtain the \emph{degraded} Gaussian BC with confidential messages (BCC).
In this section we put together some of the observations regarding the Gaussian wiretap channel and the Gaussian BC given in the previous sections so as to conclude regarding the behavior of the relevant MMSE quantities of the Gaussian BCC.

Consider a scalar additive Gaussian channel,
over which length-$\dim$ codewords are being transmitted. This is depicted as follows:
\begin{align}
\rvec{y}_{\dim} & = \sqrt{\snr_y} \rvec{x}_{\dim} + \rvec{n_1}_{\dim} \nonumber \\
\rvec{z}_{\dim} & = \sqrt{\snr_z} \rvec{x}_{\dim} + \rvec{n_2}_{\dim}
\end{align}
where $\snr_z < \snr_y$. $\rvec{n_1}_{\dim}$ and $\rvec{n_2}_{\dim}$ are standard additive Gaussian noise vectors. As in the Gaussian wiretap channel, considered in Section \ref{sec:Wiretap}, and in the Gaussian BC, considered in Section \ref{sec:GaussianBC}, also in this model we may assume that $\rvec{n_1}_{\dim} = \rvec{n_1}_{\dim} = \rvec{n}_{\dim}$ without loss of generality \cite{CsiszarKorner}. 

An $(\rate_y, \rate_z, d)$ code sequence for this channel must reliably transmit the message, $W_y$, of cardinality $2^{\dim \rate_y}$, to $\rvec{y}$ while guaranteeing an equivocation rate of $d$ and also reliably transmit the message, $W_z$, of cardinality $2^{\dim \rate_z}$, to $\rvec{z}$. These requirements can be written as follows:
\begin{align} \label{eq:wiretapDefinition}
\lim_{\dim \to \infty} \frac{1}{\dim} \Igen{W_y}{\rvec{y}_{\dim}} & = \rate_y \nonumber \\ 
\lim_{\dim \to \infty} \frac{1}{\dim} H(W_y | \rvec{z}_{\dim}) & = d \nonumber \\
\lim_{\dim \to \infty} \frac{1}{\dim} \Igen{W_z}{\rvec{z}_{\dim}} & = \rate_z.
\end{align}
The basic requirement from $(\rate_y,\rate_z, d)$ is that $\rate_y \geq d$.

The capacity region of the channel is well known \cite[Theorem 1]{CsiszarKorner}:
\begin{align}
C = \left\{ \begin{array}{l} (\rate_y, \rate_z, d): \\
0 \leq \rate_y + \rate_z \leq \Icond{\resc{x}}{\resc{y}}{\resc{u}} + \Igen{\resc{u}}{\resc{z}} \\
0 \leq d \leq \rate_y \\
d \leq \left[ \Icond{\resc{x}}{\resc{y}}{\resc{u}} - \Icond{\resc{x}}{\resc{z}}{\resc{u}} \right] \\
0 \leq \rate_z \leq \Igen{\resc{u}}{\resc{z}}
\end{array}
\right. .
\end{align}
where $\resc{u}$ is an auxiliary random variable such that $\resc{u} - \resc{x} - \resc{y} - \resc{z}$ forms a Markov chain. The capacity region of the Gaussian setting was considered in \cite{BCCfading}.

Finally, note that an optimally secure code sequence follows the same definition as given in Definition \ref{dfn:optimallySecure}; however since reliable decoding of $W_z$ is possible at $\rvec{z}$ this is equivalent also to
\begin{align}
d_{opt} = \Icond{\rvec{x}}{\rvec{y}}{W_z} - \Icond{\rvec{x}}{\rvec{z}}{W_z}.
\end{align}

\subsection{Main Results} \label{ssec:BCCmainResults}
Putting together Theorem \ref{thm:specificRates} and Theorem \ref{thm:badCodeGaussianBC} we have the following result:
\begin{cor} \label{cor:BCC_nonOptimalRate}
A code sequence transmitting $(W_y,W_z)$ at rates $(\rate_y, \rate_z)$ such that $W_y$ is reliably decoded by $\rvec{y}$ and completely secure from $\rvec{z}$, and $W_z$ is reliably decoded by
$\rvec{z}$ if and only if
\begin{align}
\MMSE( \rvec{x} ; \gamma) & = \left\{ \begin{array}{ll} \MMSE( \rvec{x} ; \gamma | W_y), & \gamma \in [0, \snr_z) \\ \MMSE( \rvec{x} ; \gamma | W_z), & \gamma \in [\snr_z, \snr_y) 
\end{array} \right. \textrm{ and} \\
\MMSE( \rvec{x} ; \gamma) & = \MMSE( \rvec{x} ; \gamma | W_y) = \MMSE( \rvec{x} ; \gamma | W_z), \quad \forall \gamma \geq \snr_y.
\end{align}
When $d < \rate_y$ (no longer complete secrecy) $\MMSE( \rvec{x} ; \gamma)$ will no longer be equal to $\MMSE( \rvec{x} ; \gamma | W_y)$ in the region of $[0, \snr_z)$.
\end{cor}
Note that we can also apply the result of Theorem \ref{thm:optimallySecureCode} to the above setting, meaning that if we require optimal security we have that
\begin{align}
\MMSE( \rvec{x} ; \gamma | W_y) = 0, \quad \forall \gamma \geq \snr_z \nonumber
\end{align}

The above corollary considers any code sequence with complete secrecy. Theorem \ref{thm:optimallySecureCode} for the Gaussian wiretap channel considers any code sequence with optimal security, and the combination of the two gives us the set of MMSE properties of code sequences that are both completely secure and optimally secure. Considering a more specific depiction of the MMSE quantities of ``good'' sequences we distinguish between two families: secrecy capacity sequences (which achieve both complete secrecy and  optimal security), and optimally secure ``good'' sequences. In both cases we aim at the highest level of equivocation and try to maximize the rate pair; however in the first family we further restrict the rate $\rate_y$ to complete secrecy ($\rate_y = d$). Next we will show that both families have the same $\MMSE( \rvec{x} ; \gamma)$ behavior, and that the second family is a subset of the family of ``good'' code sequences for the Gaussian BC.

\begin{thm} \label{thm:BCC_ps_optimal}
Consider a code sequence transmitting $(W_y,W_z)$ at rates $(\rate_y, \rate_z)$ such that $W_y$ is reliably decoded by $\rvec{y}$ and completely secure from $\rvec{z}$, and $W_z$ is reliably decoded by
$\rvec{z}$. Any ``good'' sequence has the following rate pair:
\begin{align}
(\rate_y, \rate_z) = \left( \frac{1}{2} \log \left( \frac{1 + \beta \snr_y}{1 + \beta \snr_z} \right), \frac{1}{2} \log \left( \frac{1 + \snr_z}{1 + \beta \snr_z} \right) \right)
\end{align}
for some $\beta \in [0, 1]$,
and has the following behavior:
\begin{align}
\MMSE( \rvec{x} ; \gamma) = \left\{ \begin{array}{ll} \frac{1}{1 + \gamma}, & \gamma \in [0, \snr_z) \\ \frac{\beta}{1 + \beta \gamma}, & \gamma \in [\snr_z, \snr_y) \\ 0, & \gamma \geq \snr_y \end{array} \right.,
\end{align}
\begin{align}
\MMSE( \rvec{x} ; \gamma | W_z) = \left\{ \begin{array}{ll} \frac{\beta}{1 + \beta \gamma}, & \gamma \in [0, \snr_y) \\ 0, & \gamma \geq \snr_y \end{array} \right.
\end{align}
and
\begin{align}
\MMSE( \rvec{x} ; \gamma | W_y) = \left\{ \begin{array}{ll} \frac{1}{1 + \gamma}, & \gamma \in [0, \snr_z) \\ 0 & \gamma \geq \snr_z \end{array} \right. .
\end{align}
The above is attained by a superpositions code sequence, $\rvec{x} = \rvec{v} + \rvec{u}$, where $\rvec{v}$ is an optimal Gaussian code of power $1 - \beta$ and rate $\rate_z$ (attained by considering $\rvec{u}$ as additional Gaussian noise). $\rvec{u}$, independent of $\rvec{v}$, has power $\beta$, and is a completely secure Gaussian codebook to $\rvec{y}$ with eavesdropper $\rvec{z}$, and thus attain $\rate_y$.
\end{thm}

\begin{thm} \label{thm:BCC_optimal}
Consider a code sequence transmitting $(W_y,W_z)$ at rates $(\rate_y, \rate_z)$ such that $W_y$ is reliably decoded by $\rvec{y}$ and optimally secure from $\rvec{z}$, and $W_z$ is reliably decoded by
$\rvec{z}$. Any ``good'' sequence has the following rate pair:
\begin{align}
(\rate_y, \rate_z) = \left( \frac{1}{2} \log \left( 1 + \beta \snr_y \right), \frac{1}{2} \log \left( \frac{1 + \snr_z}{1 + \beta \snr_z} \right) \right)
\end{align}
for some $\beta \in [0, 1]$, and
has the following behavior:
\begin{align} \label{thm:eq:BCC_optimallySecureMMSE}
\MMSE( \rvec{x} ; \gamma) = \left\{ \begin{array}{ll} \frac{1}{1 + \gamma}, & \gamma \in [0, \snr_z) \\ \frac{\beta}{1 + \beta \gamma}, & \gamma \in [\snr_z, \snr_y) \\ 0, & \gamma \geq \snr_y \end{array} \right.
\end{align}
and differs from the relations in Corollary \ref{cor:BCC_nonOptimalRate} only in the fact that
\begin{align}
\MMSE( \rvec{x} ; \gamma | W_y) \leq \MMSE( \rvec{x} ; \gamma), \quad \gamma \in [0, \snr_z)
\end{align}
and
\begin{align} \label{eq:BCC:thm:optimallySecureMMSEexp}
\frac{1}{2} \log \left( \frac{1 + \snr_z}{1 + \beta \snr_z} \right) = \frac{1}{2} \int_0^{\snr_z} \MMSE( \rvec{x} ; \gamma | W_y) \d \gamma .
\end{align}
The above is attained by superposition, $\rvec{x} = \rvec{v} + \rvec{u}$, where $\rvec{v}$ is an optimal Gaussian code power $1-\beta$ and rate $\rate_z$. $\rvec{u}$, independent of $\rvec{v}$, is a Gaussian optimally secure (maximum level of equivocation) code sequence for the wiretap channel to $\rvec{y}$ with eavesdropper $\rvec{z}$.
\end{thm}

\subsection{Proofs for the Gaussian BCC} \label{ssec:proofsBCC}
\begin{IEEEproof}[Proof of Theorem \ref{thm:BCC_ps_optimal}]
We consider complete secrecy, $\rate_y = d$ and wish to examine a capacity rate pair.
The rates can be written as follows:
\begin{align}
\rate_y & = \Igen{W_y}{\rvec{y}} = \Igen{\rvec{x}}{\rvec{y}} - \Icond{\rvec{x}}{\rvec{y}}{W_y} \nonumber \\
& = \frac{1}{2} \int_0^{\snr_y} \left[ \MMSE( \rvec{x} ; \gamma) - \MMSE( \rvec{x} ; \gamma | W_y) \right] \d \gamma \nonumber \\
& = \frac{1}{2} \int_{\snr_z}^{\snr_y} \left[ \MMSE( \rvec{x} ; \gamma | W_z) - \MMSE( \rvec{x} ; \gamma | W_y) \right] \d \gamma
\end{align}
where in the last transition we have taken into account the fact that this is a complete secrecy code sequence (by taking $\MMSE( \rvec{x} ; \gamma) = \MMSE( \rvec{x} ; \gamma | W_y)$ for all $\gamma \in [0, \snr_z)$) and reliable decoding of $W_z$ at $\snr_z$ (by taking $\MMSE(\rvec{x}|\gamma) = \MMSE(\rvec{x}; \gamma | W_z)$ for all $\gamma \geq \snr_z$) as given in Corollary \ref{cor:BCC_nonOptimalRate}). For the weaker receiver we have the following rate expression:
\begin{align}
\rate_z & = \Igen{W_z}{\rvec{z}} = \Igen{\rvec{x}}{\rvec{z}} - \Icond{\rvec{x}}{\rvec{z}}{W_z} \nonumber \\
& = \frac{1}{2} \int_0^{\snr_z} \left[ \MMSE( \rvec{x} ; \gamma) - \MMSE( \rvec{x} ; \gamma | W_z) \right] \d \gamma \nonumber \\
& = \frac{1}{2} \int_0^{\snr_z} \left[ \MMSE( \rvec{x} ; \gamma | W_y) - \MMSE( \rvec{x} ; \gamma | W_z) \right] \d \gamma.
\end{align}
In order to maximize both rates, we can first take $\MMSE( \rvec{x} ; \gamma | W_y) = 0$ for all $\gamma \geq \snr_z$ which is exactly the case of optimal security (since we assume reliable decoding of $W_y$ at $\snr_y$, Theorem \ref{thm:optimallySecureCode}). We thus need to maximize the following pair:
\begin{align}
\rate_y & = \frac{1}{2} \int_{\snr_z}^{\snr_y} \MMSE( \rvec{x} ; \gamma) \d \gamma = \frac{1}{2} \int_{\snr_z}^{\snr_y} \MMSE( \rvec{x} ; \gamma | W_z) \d \gamma \nonumber \\
\rate_z & = \frac{1}{2} \int_0^{\snr_z} \left[ \MMSE( \rvec{x} ; \gamma | W_y) - \MMSE( \rvec{x} ; \gamma | W_z) \right] \d \gamma.
\end{align}
Now, given that $\rate_y = \frac{1}{2} \log \left( \frac{1 + \beta \snr_y}{1 + \beta \snr_z} \right)$ for some $\beta \in [0, 1]$, we can conclude from the ``single crossing point'' property, given in Theorem \ref{thm:ScalarUniqueCrossingPoint}, that
\begin{align}
\MMSE( \rvec{x} ; \gamma | W_z) \big|_{\gamma = \snr_z} \geq \frac{\beta}{1 + \beta \snr_z} \nonumber
\end{align}
and thus,
\begin{align}
\rate_z \leq \frac{1}{2} \log(1 + \snr_z) - \frac{1}{2} \log (1 + \beta \snr_z). \nonumber
\end{align}
This is a converse proof. We first claim that this is indeed an achievable pair by using superposition. Consider codewords $\rvec{x} = \rvec{v} + \rvec{u}$ where $\rvec{v}$ has a power constraint of $1 - \beta$ and $\rvec{u}$, independent of $\rvec{v}$, has a power constraint of $\beta$. $\rvec{v}$ is an optimal Gaussian code sequence to $\rvec{z}$, and thus attains $\rate_z$ by considering $\rvec{u}$ as additional Gaussian noise. $\rvec{u}$ is a completely secure Gaussian codebook to $\rvec{y}$ with eavesdropper $\rvec{z}$, and thus attain $\rate_y$. Since $\rvec{y}$ can first reliably decode $\rvec{v}$ this concludes the achivability proof.

The above converse provides us with a full depiction of the MMSE quantities, since in order to attain the above rates with equality we need that
\begin{align}
\MMSE(\rvec{x} ; \gamma) = \frac{1}{1 + \gamma}, \quad \gamma \in [0, \snr_z),
\end{align}
\begin{align}
\MMSE( \rvec{x} ; \gamma) = \MMSE( \rvec{x} ; \gamma | W_y), \quad \gamma \in [0, \snr_z)
\end{align}
and
\begin{align}
\MMSE( \rvec{x} ; \gamma | W_z) = \frac{\beta}{1 + \beta \gamma}, \quad \gamma \in [0, \snr_y)
\end{align}
and, finally, that the code sequence is both completely secure and optimally secure; thus we have the set of relations given in Corollary \ref{cor:BCC_nonOptimalRate} and Theorem \ref{thm:optimallySecureCode}. This concludes the proof.
\end{IEEEproof}

\begin{IEEEproof}[Proof of Theorem \ref{thm:BCC_optimal}]
Observe that the rate pair given in the theorem is a capacity rate pair for the Gaussian BC, and thus according to Theorem \ref{thm:GaussianBC_codebook} we have the exact behavior of $\MMSE( \rvec{x} ; \gamma)$ for all $\gamma \geq 0$. This proves (\ref{thm:eq:BCC_optimallySecureMMSE}).
Due to reliable decoding of $W_z$ we have, according to Corollary \ref{cor:BCC_nonOptimalRate}, that $\MMSE( \rvec{x} ; \gamma) = \MMSE( \rvec{x} ; \gamma | W_z)$ for all $\gamma \geq \snr_z$. It remains to show that this is achievable by an optimally secure code sequence for the Gaussian BCC. Consider a Gaussian codebook of power $1-\beta$ to be reliably decoded at $\rvec{z}$ (of rate $\frac{1}{2} \log \left( \frac{1 + \snr_z}{1 + \beta \snr_z} \right)$) and superimpose on it, with the remaining power of $\beta$, an independent Gaussian optimally secure (maximum level of equivocation) code sequence for the wiretap channel to $\rvec{y}$ with eavesdropper $\rvec{z}$, meaning an
\begin{align}
(\rate_y, d) = \left( \frac{1}{2} \log(1 + \beta \snr_y), \frac{1}{2} \log \left(\frac{1 + \beta \snr_y}{1 + \beta \snr_z} \right) \right)
\end{align}
rate-equivocation pair. Thus, the rate-pair is achievable and is optimally secure:
\begin{align}
d = d_{opt} & = \Igen{\rvec{x}}{\rvec{y}} - \Igen{\rvec{x}}{\rvec{z}} \nonumber \\
& = \Icond{\rvec{x}}{\rvec{y}}{W_z} - \Icond{\rvec{x}}{\rvec{z}}{W_z} \nonumber \\
& = \frac{1}{2} \log(1 + \beta \snr_y) - \frac{1}{2} \log(1 + \beta \snr_z).
\end{align}
The only MMSE quantity that is not fully defined is $\MMSE( \rvec{x} ; \gamma | W_y)$ in the region of $[0, \snr_z)$. For this we can use $\rate_y$ and claim the following:
\begin{align}
\rate_y & = \Igen{\rvec{x}}{\rvec{y}} - \Icond{\rvec{x}}{\rvec{y}}{W_y} \nonumber \\
& = \frac{1}{2} \int_0^{\snr_y} \left[ \MMSE( \rvec{x} ; \gamma) - \MMSE( \rvec{x} ; \gamma | W_y) \right] \d \gamma \nonumber \\
& = \frac{1}{2} \int_0^{\snr_z} \frac{1}{1 + \gamma} - \MMSE( \rvec{x} ; \gamma | W_y) \d \gamma + \frac{1}{2} \log \left( \frac{ 1 + \beta \snr_y}{1 + \beta \snr_z} \right)
\end{align}
which leads to the following:
\begin{align}
\frac{1}{2} \int_0^{\snr_z} \MMSE( \rvec{x} ; \gamma | W_y) \d \gamma = \frac{1}{2} \log \left( \frac{1 + \snr_z}{1 + \beta \snr_z} \right).
\end{align}
This concludes the proof.
\end{IEEEproof}

\section{Gaussian BC and BCC with MMSE Constraints} \label{sec:Disturbance}
In the previous sections we examined the effect of the transmission on unintended receivers at other SNRs. We have done so by fully depicting the behavior of the input-output mutual information and MMSE function for all SNRs. In this section we examine the trade-off between achievable rates and MMSE constraints. These constraints come to limit the amount of disturbance the transmission has on unintended receivers at other SNRs. These results are a direct extension of \cite{IMMSEtradeoff} where the trade-off between the point-to-point rate and MMSE has been investigated. It was shown that the optimal scheme attaining both the maximal possible point-to-point rate and complying with the MMSE constraint is superposition, supporting the good performance of the Han and Kobayashi scheme \cite{HanKobayashi} in the two-user Gaussian interference channel.

We first consider an MMSE constraint limiting the capacity region of the scalar Gaussian BC, depicted in (\ref{eq:modelGaussianBC}). We then consider the effect of this constraint on the Gaussian BCC. More specifically, we consider a code sequence transmitting the message pair $(W_y, W_z)$ such that $W_y$ must be reliably decoded by $\rvec{y}$ and completely secure from $\rvec{z}$, and $W_z$ must be reliably decoded by $\rvec{z}$.

On top of these requirements we now consider an additional constraint:
\begin{align} \label{eq:constraintSNRu}
\MMSE( \rvec{x} ; \snr_u ) \leq \frac{\alpha}{1 + \alpha \snr_u}
\end{align}
for some $\alpha \in [0,1]$.
Of course, the effect of this constraints varies whether $\snr_u < \snr_z$ or $\snr_u \in (\snr_z, \snr_y)$.

\subsection{Main Results} \label{ssec:BCCdisturbanceMainResults}
The first theorem considers the effect of the constraint in (\ref{eq:constraintSNRu}) on the capacity region of the Gaussian BC.

\begin{thm} \label{thm:MMSEconstraintGaussianBC}
Given the MMSE constraint in (\ref{eq:constraintSNRu}) the set of achievable rate pairs for the \emph{degraded} Gaussian BC depicted in (\ref{eq:modelGaussianBC}) is the following:

Assuming that $\snr_u \in (\snr_z, \snr_y)$. The region is given as a union of
\begin{align}
(\rate_y, \rate_z) = \left(\frac{1}{2} \log( 1 + \beta \snr_y), \frac{1}{2} \log \left( \frac{1 + \snr_z}{1 + \beta \snr_z } \right) \right)
\end{align}
(the expression in (\ref{eq:thm:BC:capacityPoint})) for $\beta \in [0, \alpha]$  and
\begin{align}
(\rate_y, \rate_z) = \left(\frac{1}{2} \log (1 + \alpha \snr_y) + \frac{1}{2} \log \left( 1 + \frac{(\beta - \alpha) \snr_u }{1 + \alpha \snr_u} \right) 
,\frac{1}{2} \log\left( \frac{1 + \snr_z}{1 + \beta \snr_z} \right) \right)
\end{align}
for $\beta \in (\alpha, 1]$. The second region is achievable using a 3-layer superposition Gaussian code sequence. The message to be decoded from $\rvec{z}$ has $1 - \beta$ of the power. The second layer can already be reliably decoded at $\snr_u$ and has power $\beta - \alpha$, and the final layer to be decoded from $\rvec{y}$ has power $\alpha$.

Assuming that $\snr_u \in [0, \snr_z)$ the capacity is again a union of two regions. The first is
\begin{align}
(\rate_y, \rate_z) = \left( \frac{1}{2} \log (1 + \alpha \snr_y) + \lambda \frac{1}{2} \log \left( \frac{1 + \snr_u}{1 + \alpha \snr_u} \right) ,
(1- \lambda) \frac{1}{2} \log \left( \frac{1 + \snr_u}{1 + \alpha \snr_u} \right)\right)
\end{align}
where $\lambda \in [0,1]$. This region is obtained by time sharing. In both schemes we use a two-layer Gaussian superposition code sequence where the first layer, of power $1 - \alpha$, can be reliably decoded at $\snr_u$.

The second region is
\begin{align} \label{eq:snr_u_middle_superpositionBC}
(\rate_y, \rate_z) = \left( \frac{1}{2} \log(1 + \beta \snr_y),
\frac{1}{2} \log \left( \frac{1 + \snr_u}{1 + \alpha \snr_u} \frac{1 + \alpha \snr_z}{1 + \beta \snr_z} \right) \right)
\end{align}
where $\beta \in [0, \alpha)$. This region can be obtained by a 3-layer Gaussian superposition code sequence. The first layer, of power $1- \alpha$, can be reliably decoded at $\snr_u$. The second layer, of power $\alpha - \beta$, can be reliably decoded by $\rvec{z}$. The third layer, of power $\beta$, is decoded from $\rvec{y}$.
\end{thm}

\begin{figure}[h]
\begin{center}
\setlength{\unitlength}{.1cm}
    \includegraphics[width=1.1\textwidth]{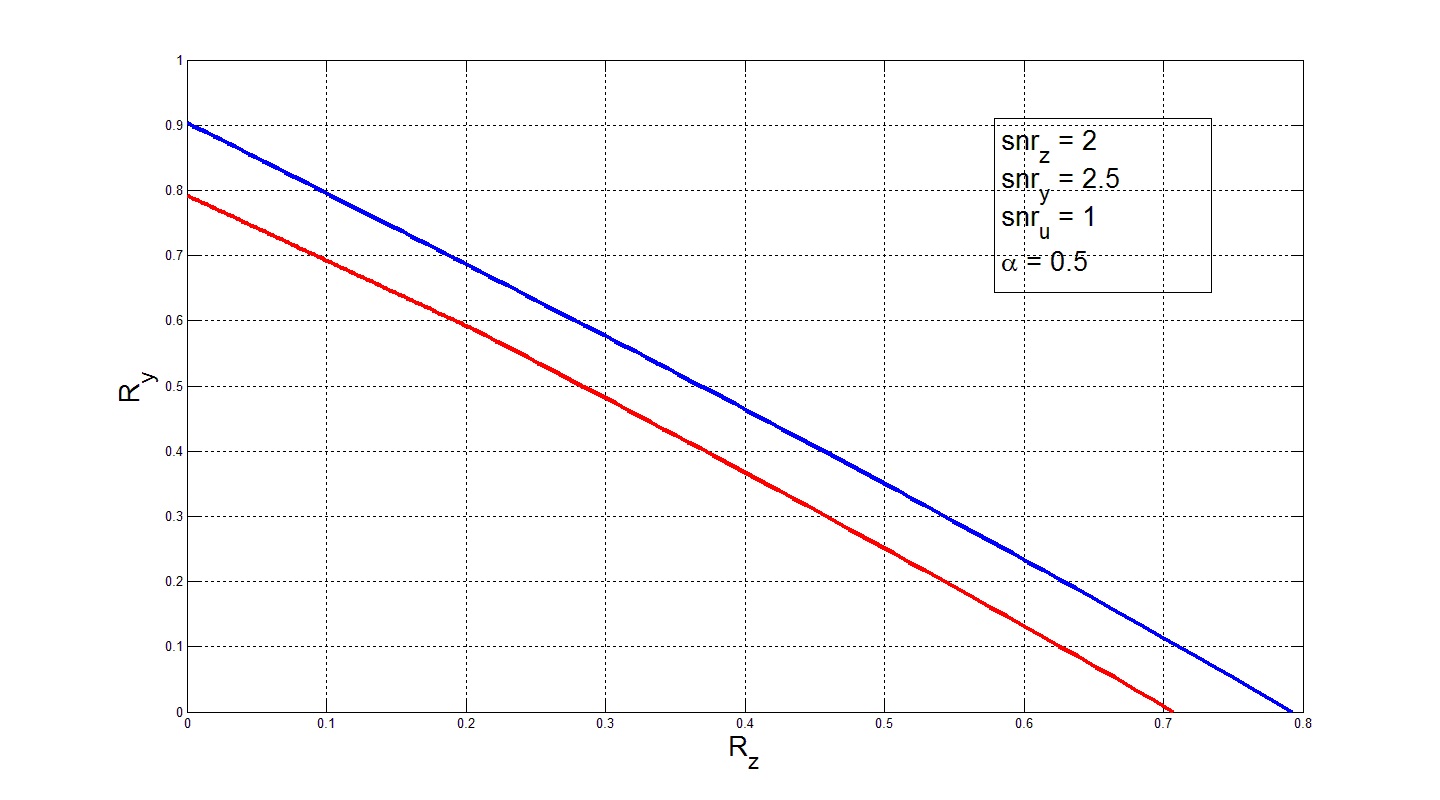}
    \caption{In the above figure we compare the capacity region of the Gaussian BC with no MMSE constraint (in blue) with the capacity region given the MMSE constraint (in red). In this example $\snr_u \in [0, \snr_z)$.}
 \label{figure:BC_MMSEconstraint1}
\end{center}
\end{figure}

\begin{figure}[h]
\begin{center}
\setlength{\unitlength}{.1cm}
    \includegraphics[width=1.1\textwidth]{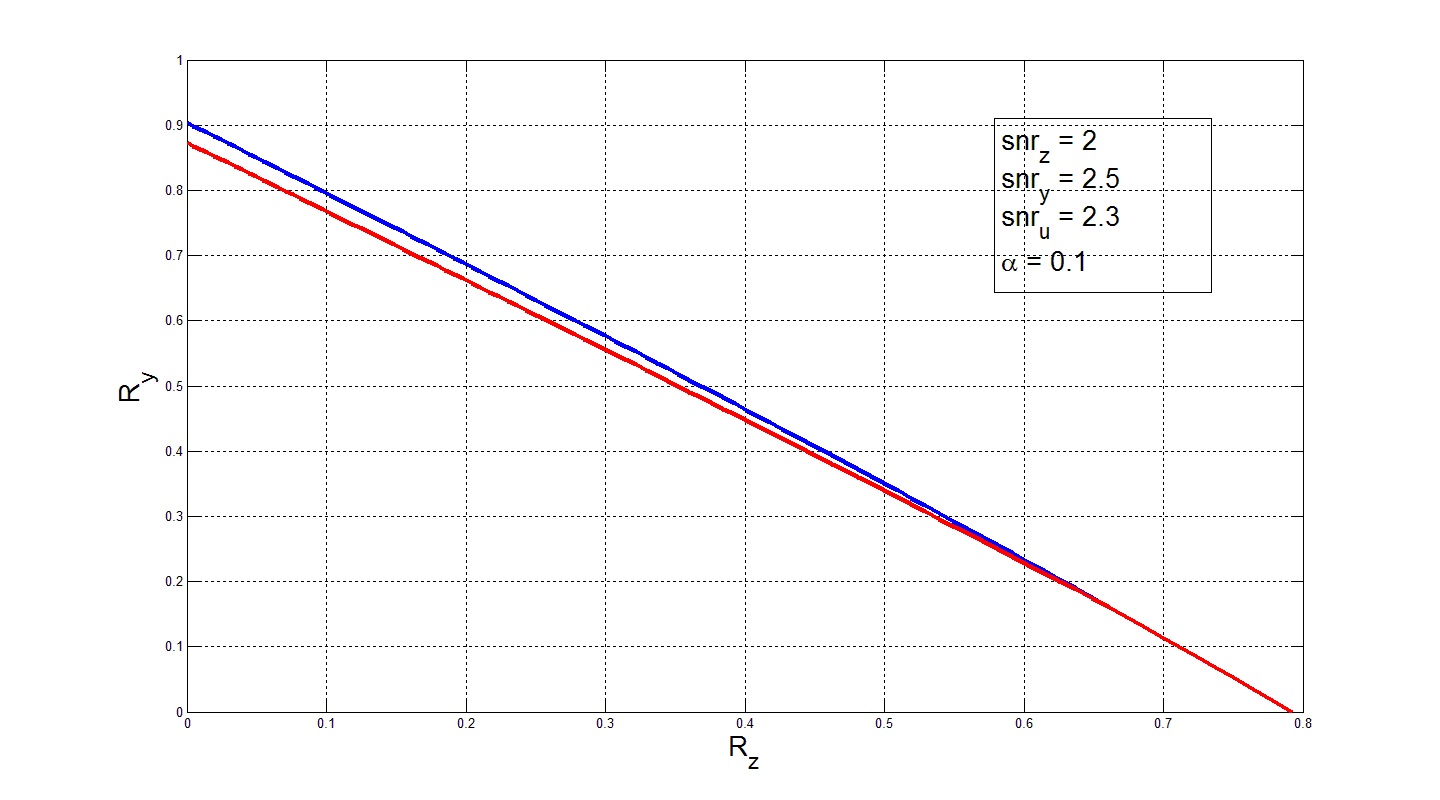}
    \caption{In the above figure we compare the capacity region of the Gaussian BC with no MMSE constraint (in blue) with the capacity region given the MMSE constraint (in red). In this example $\snr_u \in (\snr_z, \snr_y)$.}
 \label{figure:BC_MMSEconstraint2}
\end{center}
\end{figure}

An example for the above result is depicted in Figures \ref{figure:BC_MMSEconstraint1} and \ref{figure:BC_MMSEconstraint2}, where we show the reduction of the capacity region as compared to the case with no MMSE constraint.

\begin{rem}
The above regions of rate pairs of reliable communication over the Gaussian BC given an MMSE constraint at some lower SNR reduce to the result of \cite[Theorem 3]{IMMSEtradeoff} where given such a constraint the goal was to maximize the point-to-point reliable rate. When $\snr_u \in (\snr_z, \snr_y)$ this can be seen by enforcing $\rate_z = 0$ in which case we obtain a two-layer superposition code to $\rvec{y}$ with a rate-splitting coefficient of $\alpha$. When $\snr_u \in [0, \snr_z)$ we can either take $\lambda = 1$ in the time sharing argument, or alternatively take $\beta = \alpha$ in (\ref{eq:snr_u_middle_superpositionBC}) in which case we maximize the sum-rate, $\rate_y + \rate_z$. In both cases we again obtain the same expression which can be obtained by a two-layer Gaussian superposition code with rate splitting coefficient of $\alpha$.
\end{rem}

We now provide our main results on the BCC with an MMSE disturbance constraints. We consider a code pair sequence transmitting the message pair $(W_y, W_z)$ such that $W_y$ must be reliably decoded by $\rvec{y}$ and completely secure from $\rvec{z}$, and $W_z$ must be reliably decoded by $\rvec{z}$.
For these requirements Corollary \ref{cor:BCC_nonOptimalRate} defines the properties of the MMSE functions. Moreover, from the proof we see that the optimal rate-pair is obtained when we further assume optimal security.  
The MMSE constraint (\ref{eq:constraintSNRu}) at $\snr_u$ requires us to distinguish between the two cases of $\snr_u \in [0, \snr_z)$ and $\snr_u \in (\snr_z, \snr_y)$ and consider them separately in the next two theorems.

\begin{thm} \label{thm:SNRu_between}
Consider a code pair sequence transmitting over the Gaussian channel the message pair $(W_y, W_z)$ such that $W_y$ is reliably decoded by $\rvec{y}$ and completely secure from $\rvec{z}$, and $W_z$ is reliably decoded by $\rvec{z}$. In addition we have the MMSE constraint (\ref{eq:constraintSNRu}) at $\snr_u \in (\snr_z, \snr_y)$. The maximum rate pair is of the following form:
\begin{align}
(\rate_y, \rate_z) = \left( \frac{1}{2} \log \left( \frac{1 + \beta \snr_u}{1 + \beta \snr_z} \right) + \frac{1}{2} \log \left( \frac{1 + \alpha \snr_y}{1 + \alpha \snr_u} \right) ,\frac{1}{2} \log\left( \frac{1 + \snr_z}{1 + \beta \snr_z} \right)\right)
\end{align}
for some $\beta \in [\alpha, 1]$. These rate pairs are obtained by a 3-layer superposition code sequence, where the first layer is an optimal Gaussian code sequence to $\rvec{z}$, of power $1 - \beta$, the second is a secrecy capacity Gaussian code sequence to the unintended receiver at $\snr_u$ with eavesdropper $\rvec{z}$, of power $\beta - \alpha$, and the last layer is a secrecy capacity Gaussian code sequence to $\rvec{y}$ with eavesdropper $\rvec{z}$, of power $\alpha$. These rates have the following behavior:
\begin{align}
\MMSE( \rvec{x} ; \gamma) = \left\{ \begin{array}{ll} \frac{1}{1 + \gamma}, & \gamma \in [0, \snr_z) \\ \frac{\beta}{1 + \beta \gamma}, & \gamma \in [\snr_z, \snr_u) \\ \frac{\alpha}{1 + \alpha \gamma}, & \gamma \in [\snr_u, \snr_y) \\ 0, &\gamma \geq \snr_y \end{array} \right.
\end{align}
and
\begin{align}
\MMSE( \rvec{x} ; \gamma | W_z) = \left\{ \begin{array}{ll} \frac{\beta}{1 + \beta \gamma}, & \gamma \in [0, \snr_u) \\ \frac{\alpha}{1 + \alpha \gamma}, & \gamma \in [\snr_u, \snr_y) \\ 0, & \gamma \geq \snr_y \end{array} \right.
\end{align}
and $\MMSE( \rvec{x} ; \gamma | W_y)$ is as in Theorem \ref{thm:BCC_ps_optimal}. \\
In addition the rate pairs depicted in Theorem \ref{thm:BCC_ps_optimal} for $\beta \in [0,\alpha)$ and their MMSE behavior are also optimal and trivially comply with the constraint.
\end{thm}

\begin{thm} \label{thm:SNRu_low}
Consider a code pair sequence transmitting over the Gaussian channel the message pair $(W_y, W_z)$ such that $W_y$ is reliably decoded by $\rvec{y}$ and completely secure from $\rvec{z}$, and $W_z$ is reliably decoded by $\rvec{z}$. In addition we have the MMSE constraint (\ref{eq:constraintSNRu}) at $\snr_u \in [0, \snr_z)$. The maximum rate pair is of the following form:
\begin{align}
(\rate_y, \rate_z) = \left( \frac{1}{2} \log \left( \frac{1 + \beta \snr_y}{1 + \beta \snr_z} \right) ,\frac{1}{2} \log \left( 1 + \frac{(1- \alpha) \snr_u}{1 + \alpha \snr_u}\right) + \frac{1}{2} \log \left( 1 + \frac{(\alpha - \beta) \snr_z}{ 1 + \beta \snr_z}\right) \right)
\end{align}
for some $\beta \in [0, \alpha]$. These rate pairs are obtained by a 3-layer superposition code sequence. The first layer, of power $1- \alpha$, is an optimal Gaussian code sequence for the unintended receiver at $\snr_u$, the second layer is of power $\alpha - \beta$ to $\rvec{z}$, and the last layer is a secrecy capacity achieving code sequence of power $\beta$ designed for receiver $\rvec{y}$ with eavesdropper $\rvec{z}$.
These rate pairs have the following MMSE behavior:
\begin{align}
\MMSE( \rvec{x} ; \gamma) = \left\{ \begin{array}{ll} \frac{1}{1 + \gamma}, & \gamma \in [0, \snr_u) \\ \frac{\alpha}{1 + \alpha \gamma}, & \gamma \in [\snr_u, \snr_z) \\ \frac{\beta}{1 + \beta \gamma}, & \gamma \in [\snr_z, \snr_y) \\ 0, & \gamma \geq \snr_y \end{array} \right.
\end{align}
and
\begin{align}
\MMSE( \rvec{x} ; \gamma | W_y) = \left\{ \begin{array}{ll} \frac{1}{1 + \gamma}, & \gamma \in [0, \snr_u) \\ \frac{\alpha}{1 + \alpha \gamma}, & \gamma \in [\snr_u, \snr_z) \\ 0, & \gamma \geq \snr_z \end{array} \right.
\end{align}
and $\MMSE( \rvec{x} ; \gamma | W_z)$ is as depicted in Theorem \ref{thm:BCC_ps_optimal}.
\end{thm}

From the above two theorems we can clearly see the different effects the constraint can have. In Theorem \ref{thm:SNRu_between} the constraint has no effect on the rate to the weaker receiver and only the wiretap code sequence has to be altered in order to comply. In Theorem \ref{thm:SNRu_low} the situation is reversed. In both cases a simple superposition scheme maximized the rate under the constraint.

\begin{figure}[h]
\begin{center}
\setlength{\unitlength}{.1cm}
    \includegraphics[width=1.1\textwidth]{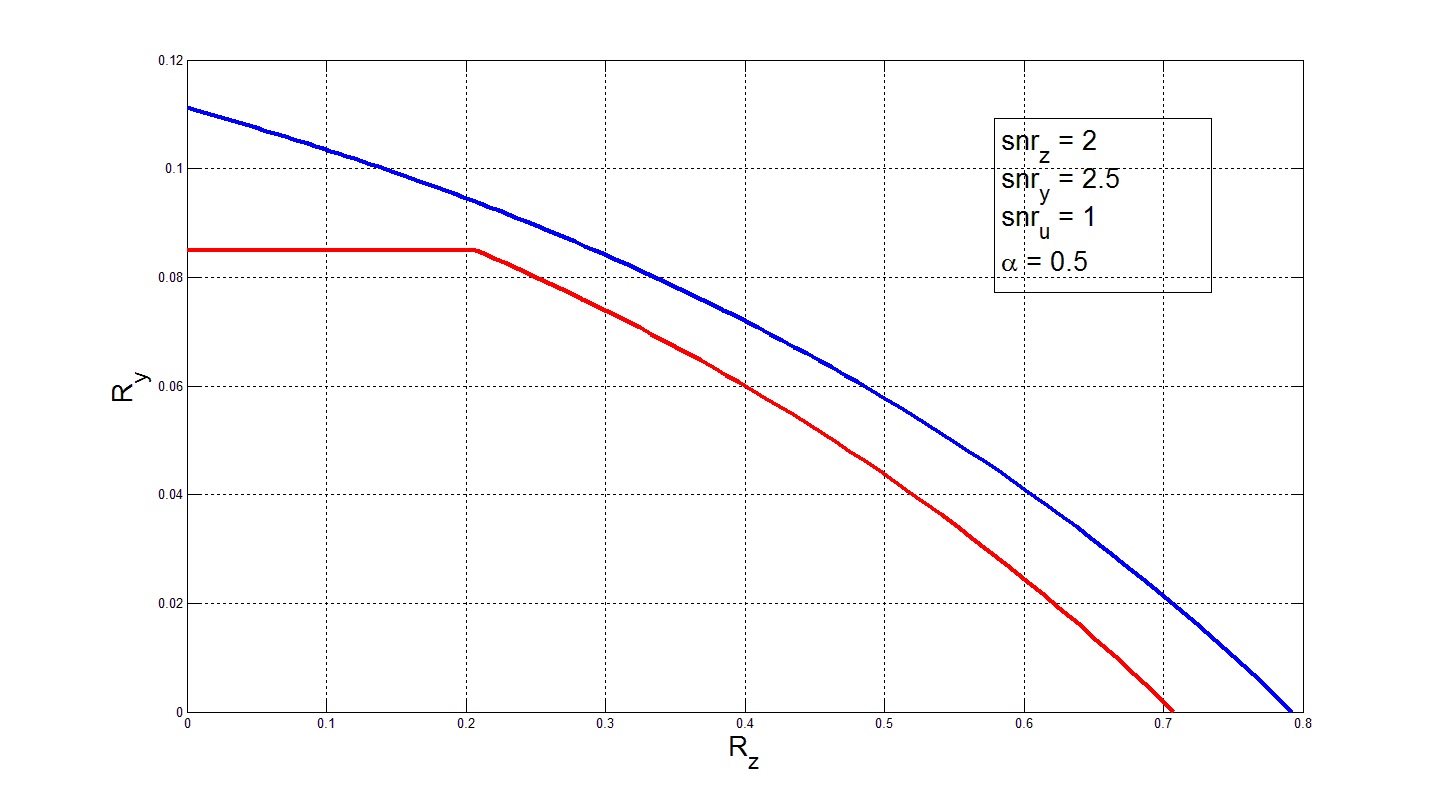}
    \caption{In the above figure we compare the capacity region of the Gaussian BCC with no MMSE constraint (in blue) with the capacity region given the MMSE constraint (in red). In this example $\snr_u \in [0, \snr_z)$.}
 \label{figure:BCC_MMSEconstraint1}
\end{center}
\end{figure}

\begin{figure}[h]
\begin{center}
\setlength{\unitlength}{.1cm}
    \includegraphics[width=1.1\textwidth]{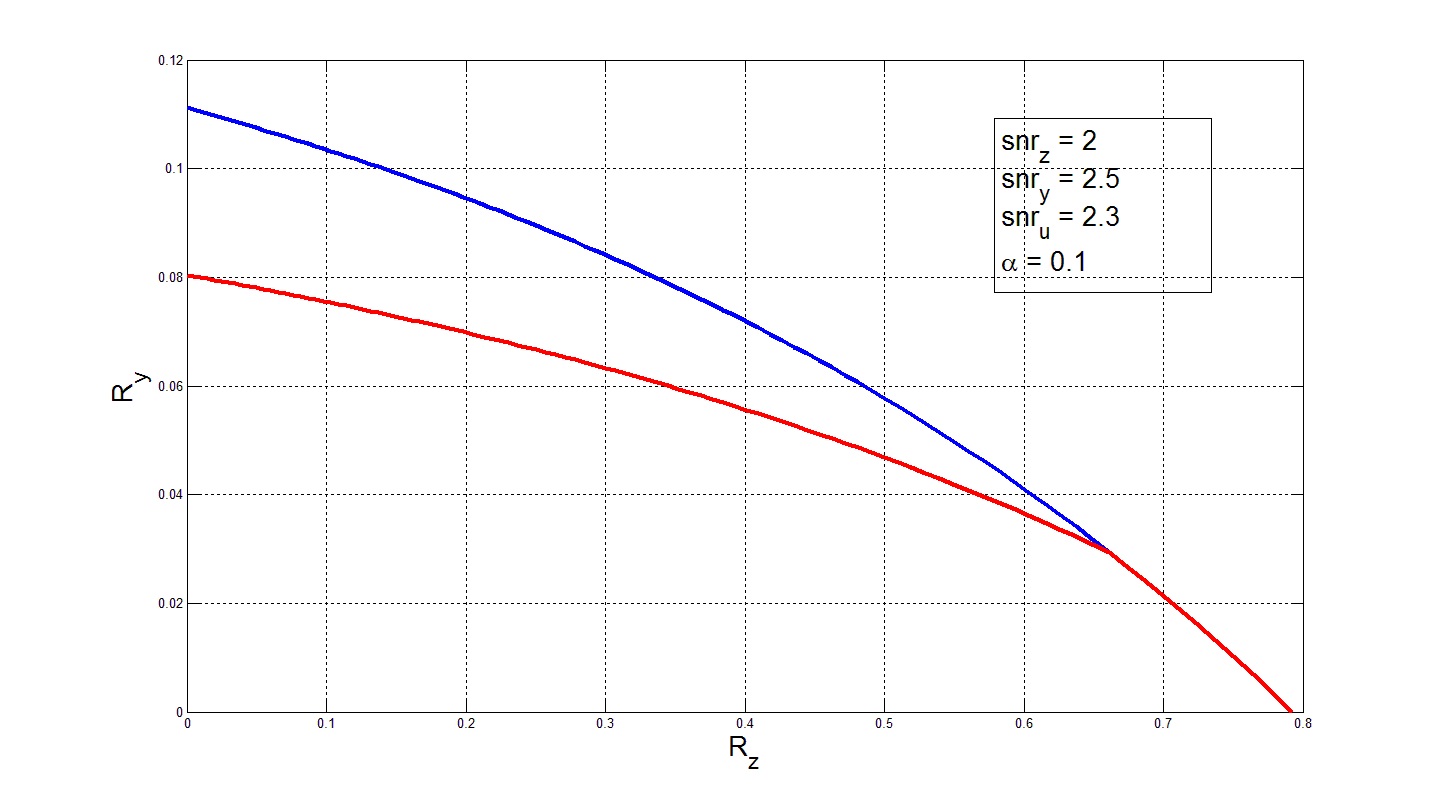}
    \caption{In the above figure we compare the capacity region of the Gaussian BCC with no MMSE constraint (in blue) with the capacity region given the MMSE constraint (in red). In this example $\snr_u \in (\snr_z, \snr_y)$.}
 \label{figure:BCC_MMSEconstraint2}
\end{center}
\end{figure}

An example for the above results is depicted in Figures \ref{figure:BCC_MMSEconstraint1} and \ref{figure:BCC_MMSEconstraint2}, where we show the reduction of the capacity region as compared to the case with no MMSE constraint.

\subsection{Proofs for the Gaussian BC and BCC with MMSE Constraints} \label{ssec:BCCDisturbanceProofs}
\begin{IEEEproof}[Proof of Theorem \ref{thm:MMSEconstraintGaussianBC}]
We assume $\snr_u \in (\snr_z, \snr_y)$.
We arbitrarily set $\rate_z$ as follows:
\begin{align}
\rate_z = \frac{1}{2} \log\left( \frac{1 + \snr_z}{1 + \beta \snr_z} \right)
\end{align}
for some $\beta \in [0,1]$.
In the proof of Theorem \ref{thm:GaussianBC_codebook} we have shown that
\begin{align}
\Icond{\rvec{x}}{\rvec{z}}{W_z} \leq \Igen{\rvec{x}}{\rvec{z}} - \rate_z.
\end{align}
From which we concluded (using Theorem \ref{thm:ScalarUniqueCrossingPoint}) that
\begin{align}
\MMSE( \rvec{x} ; \snr_z | W_z) \leq \frac{\beta}{1 + \beta \snr_z}.
\end{align}
Due to Theorem \ref{thm:badCodeGaussianBC} we can also conclude that
\begin{align}
\MMSE( \rvec{x} ; \snr_z) \leq  \frac{\beta}{1 + \beta \snr_z }.
\end{align}
Note that if $\beta \leq \alpha$ we comply with the additional MMSE constraint according to Theorem \ref{thm:ScalarUniqueCrossingPoint} and thus the region is as in (\ref{eq:thm:BC:capacityPoint}). The interesting question is what happens when $\beta \in (\alpha, 1]$. In this case we have two MMSE constraints at $\snr_u$ and $\snr_z$ and thus can use \cite[Theorem 5]{IMMSEtradeoff} and conclude that
\begin{align}
\Igen{\rvec{x}}{\rvec{y}} \leq \frac{1}{2} \log \left( \frac{1 + \snr_z}{1 + \beta \snr_z} \frac{1 + \beta \snr_u}{1 + \alpha \snr_u} \right) + \frac{1}{2} \log (1 + \alpha \snr_y).
\end{align}
Due to the data-processing inequality
\begin{align}
\rate_z + \rate_y \leq \Igen{\rvec{x}}{\rvec{y}}
\end{align}
and thus,
\begin{align}
\rate_y & \leq \frac{1}{2} \log \left( \frac{1 + \beta \snr_u}{1 + \alpha \snr_u} \right) + \frac{1}{2} \log (1 + \alpha \snr_y) \nonumber \\
& = \frac{1}{2} \log (1 + \alpha \snr_y) + \frac{1}{2} \log \left( 1 + \frac{(\beta - \alpha) \snr_u }{1 + \alpha \snr_u} \right).
\end{align}
This rate pair is attainable using a 3-layer Gaussian superposition code sequence. The message to be decoded from $\rvec{z}$ has $1 - \beta$ of the power, the second layer can already be reliably decoded at $\snr_u$ and has power of $\beta - \alpha$, and the final layer to be decoded from $\rvec{y}$ has power of $\alpha$. Each layer can be reliably decoded by fully decoding and removing previous layers and considering all subsequent layers as additive Gaussian noise.

The case of $\snr_u \leq \snr_z$ is a bit more complex. Consider first the problem of maximizing $\Igen{\rvec{x}}{\rvec{z}}$ given the MMSE constraint at $\snr_u$. This follows \cite[Theorem 3]{IMMSEtradeoff}:
\begin{align}
\Igen{\rvec{x}}{\rvec{z}} \leq \frac{1}{2} \log \left( \frac{1 + \snr_u}{1 + \alpha \snr_u} \right) + \frac{1}{2} \log ( 1 + \alpha \snr_z).
\end{align}
Thus, we consider the following arbitrary rate of transmission of $W_z$:
\begin{align}
\rate_z = \frac{1}{2} \log \left( \frac{1 + \snr_u}{1 + \alpha \snr_u} \right) + \frac{1}{2} \log ( 1 + \alpha \snr_z) - \frac{1}{2} \log (1 + \beta \snr_z).
\end{align}
When $\beta = 0$ we obtain the maximum possible rate of $W_z$.
As for the upper bound on $\beta$, we note that the rate must be non-negative, thus
\begin{align}
\beta \leq \frac{(1- \alpha)\snr_u + \alpha \snr_z(1 + \snr_u)}{\snr_z (1 + \alpha \snr_u)} \equiv \beta_{\max}.
\end{align}
Following the approach in the proof of Theorem \ref{thm:GaussianBC_codebook} we can conclude from the above setting that
\begin{align}
\MMSE( \rvec{x} ; \snr_z | W_z) \leq \frac{\beta}{1 + \beta \snr_z }.
\end{align}
Due to Theorem \ref{thm:badCodeGaussianBC} we can also conclude that
\begin{align}
\MMSE( \rvec{x} ; \snr_z) \leq  \frac{\beta}{1 + \snr_z \beta}.
\end{align}
This constraint is valid only when $\beta < \alpha$, otherwise it is redundant (as the constraint at $\snr_u$ already sets a stronger upper bound at $\snr_z$). Thus, we split the analysis to two cases. First $\beta \in [\alpha, \beta_{\max}]$. In this case we use \cite[Theorem 3]{IMMSEtradeoff}, meaning maximizing the mutual information given a single MMSE constraint at $\snr_u$:
\begin{align}
\rate_y + \rate_z & \leq \Igen{\rvec{x}}{\rvec{y}} \nonumber \\
& \leq \frac{1}{2} \log \left( \frac{1 + \snr_u}{1 + \alpha \snr_u}\right) + \frac{1}{2} \log(1 + \alpha \snr_y)
\end{align}
and thus
\begin{align}
\rate_y \leq \frac{1}{2} \log\left( \frac{1 + \alpha \snr_y}{1 + \alpha \snr_z} \right) + \frac{1}{2} \log (1 + \beta \snr_z).
\end{align}
When $\beta = \beta_{\max}$ we have the following rate pair:
\begin{align}
(\rate_y, \rate_z) = \left( \frac{1}{2} \log \left( \frac{1 + \snr_u}{1 + \alpha \snr_u} \right) + \frac{1}{2} \log (1 + \alpha \snr_y) ,0 \right)
\end{align}
and is clearly achievable using the Gaussian superposition codebook as shown in \cite[Theorem 3]{IMMSEtradeoff} ($\rate_z = 0$). If $\beta = \alpha$ we have
\begin{align}
(\rate_y, \rate_z) = \left( \frac{1}{2} \log (1 + \alpha \snr_y) ,\frac{1}{2} \log \left( \frac{1 + \snr_u}{1 + \alpha \snr_u} \right) \right).
\end{align}
This is again achievable using a Gaussian superposition codebook sequence, where the first layer is of power $1-\alpha$ and can be reliably decoded by a user at SNR of $\snr_u$ (and thus also by $\rvec{z}$), while considering the second layer as additive Gaussian noise. The second layer is of power $\alpha$, and can be reliably decoded from $\rvec{y}$, once the first layer is reliably decoded and removed.

Now, the rate pair we obtained:
\begin{align}
(\rate_y, \rate_z) = \left( \frac{1}{2} \log (1 + \alpha \snr_y) + \frac{1}{2} \log \left( \frac{1 + \beta \snr_z}{1 + \alpha \snr_z} \right) , 
\frac{1}{2} \log \left( \frac{1 + \snr_u}{1 + \alpha \snr_u} \right) - \frac{1}{2} \log \left( \frac{1 + \beta \snr_z}{1 + \alpha \snr_z} \right) \right)
\end{align}
which given that $\beta \in [\alpha, \beta_{\max} ]$ is equivalent to
\begin{align}
(\rate_y, \rate_z) = \left( \frac{1}{2} \log (1 + \alpha \snr_y) + \lambda \frac{1}{2} \log \left( \frac{1 + \snr_u}{1 + \alpha \snr_u} \right) , 
(1- \lambda) \frac{1}{2} \log \left( \frac{1 + \snr_u}{1 + \alpha \snr_u} \right)\right)
\end{align}
meaning that every point in this region can be obtained by time-sharing between the above two extreme cases of $\beta = \beta_{\max}$ and $\beta = \alpha$.

Second, consider the case of $\beta \in [0, \alpha)$. In this case we use \cite[Theorem 5]{IMMSEtradeoff}, meaning maximizing the mutual information given two MMSE constraint at $\snr_u$ and $\snr_z$:
\begin{align}
\rate_y + \rate_z & \leq \Igen{\rvec{x}}{\rvec{y}}  \\
& \leq \frac{1}{2} \log \left( \frac{1 + \snr_u}{1 + \alpha \snr_u} \frac{1 + \alpha \snr_z}{1 + \beta \snr_z} \right) + \frac{1}{2} \log(1 + \beta \snr_y)\nonumber
\end{align}
and thus
\begin{align}
\rate_y \leq \frac{1}{2} \log(1 + \beta \snr_y).
\end{align}
The rate pair is then given by
\begin{align}
(\rate_y, \rate_z) = \left( \frac{1}{2} \log(1 + \beta \snr_y) , \frac{1}{2} \log \left( \frac{1 + \snr_u}{1 + \alpha \snr_u} \frac{1 + \alpha \snr_z}{1 + \beta \snr_z} \right) \right)
\end{align}
which can be obtained by a 3-layer Gaussian superposition code sequence: the first layer of power $1- \alpha$, can be reliably decoded at $\snr_u$ while considering all subsequence layers as additive Gaussian noise. The second layer of power $\alpha - \beta$ can be reliably decoded by $\rvec{z}$ after reliably decoding the first layer and considering the third layer as additive Gaussian noise. The third layer of power $\beta$ is decoded after reliably decoding and removing all previous layers.
\end{IEEEproof}

\begin{IEEEproof}[Proof of Theorem \ref{thm:SNRu_between}]
First, for $\beta \in [0, \alpha)$, from Theorem \ref{thm:BCC_ps_optimal} we can see that an optimal rate pair complies with the constraint. Thus, we can focus only on $\beta \in [\alpha, 1]$.
As we consider complete secrecy, given the MMSE properties of Corollary \ref{cor:BCC_nonOptimalRate} we have the following expression for the rate pair:
\begin{align}
\rate_y & = \frac{1}{2} \int_{\snr_z}^{\snr_y} \left[ \MMSE( \rvec{x} ; \gamma) - \MMSE( \rvec{x}; \gamma | W_y) \right] \d \gamma \nonumber \\
 & = \frac{1}{2} \int_{\snr_z}^{\snr_y} \left[ \MMSE( \rvec{x} ; \gamma | W_z) - \MMSE( \rvec{x}; \gamma | W_y) \right] \d \gamma \nonumber \\
 & \leq \frac{1}{2} \int_{\snr_z}^{\snr_y}  \MMSE( \rvec{x} ; \gamma | W_z) \d \gamma \nonumber \\
\rate_z & = \frac{1}{2} \int_0^{\snr_z} \left[ \MMSE(\rvec{x} ; \gamma) - \MMSE(\rvec{x} ; \gamma | W_z) \right] \d \gamma
\end{align}
where the inequality is obtained when $\MMSE( \rvec{x}; \gamma | W_y) = 0$ for all $\gamma \geq \snr_z$, that is, an optimally secure code sequence.
Recall that (Corollary \ref{cor:BCC_nonOptimalRate})
\begin{align}
\MMSE(\rvec{x} ; \gamma) = \MMSE(\rvec{x} ; \gamma | W_z), \quad \forall \gamma \geq \snr_z,
\end{align}
and thus the constraint at $\snr_u \in (\snr_z, \snr_y)$ is also a constraint on $\MMSE(\rvec{x} ; \gamma | W_z)$. Consider a function $\MMSE(\rvec{x} ; \gamma | W_z)$ that satisfies the constraint and define $\beta$ such that
\begin{align} \label{eq:proofBCCdefineBeta}
\frac{1}{2} \int_0^{\snr_z} \MMSE(\rvec{x} ; \gamma | W_z) \d \gamma = \frac{1}{2} \log (1 + \beta \snr_z)
\end{align}
where $\beta \in [\alpha, 1]$. By applying the ``single crossing property'', given in Theorem \ref{thm:ScalarUniqueCrossingPoint}, twice, first on (\ref{eq:proofBCCdefineBeta}) and second on the constraint $\MMSE(\rvec{x} ; \snr_u | W_z) \leq \frac{\alpha}{1 + \alpha \snr_u}$ we obtain the following upper bounds on the rates:
\begin{align}
\rate_y & \leq \frac{1}{2} \int_{\snr_z}^{\snr_y} \MMSE( \rvec{x} ; \gamma | W_z) \d \gamma \nonumber \\
& = \frac{1}{2} \int_{\snr_z}^{\snr_u} \MMSE( \rvec{x} ; \gamma | W_z) \d \gamma + \frac{1}{2} \int_{\snr_u}^{\snr_y} \MMSE( \rvec{x} ; \gamma | W_z) \d \gamma \nonumber \\
& \leq \frac{1}{2} \int_{\snr_z}^{\snr_u} \frac{\beta}{1 +  \beta \gamma} \d \gamma + \frac{1}{2} \int_{\snr_u}^{\snr_y} \MMSE( \rvec{x} ; \gamma | W_z) \d \gamma \nonumber \\
& \leq \frac{1}{2} \int_{\snr_z}^{\snr_u} \frac{\beta}{1 +  \beta \gamma} \d \gamma + \frac{1}{2} \int_{\snr_u}^{\snr_y} \frac{\alpha}{1 + \alpha \gamma} \d \gamma \nonumber \\
& = \frac{1}{2} \log( 1 + \beta \snr_u) - \frac{1}{2} \log( 1 + \beta \snr_z) + \frac{1}{2} \log( 1 + \alpha \snr_y) - \frac{1}{2} \log( 1+ \alpha \snr_u).
\end{align}
We also have
\begin{align}
\rate_z & = \frac{1}{2} \int_0^{\snr_z} \left[ \MMSE(\rvec{x} ; \gamma) - \MMSE(\rvec{x} ; \gamma | W_z) \right] \d \gamma \nonumber \\
& \leq \frac{1}{2} \log(1 + \snr_z ) - \frac{1}{2} \log( 1 + \beta \snr_z).
\end{align}
The achievability of the above rate pair follows simply by using a superposition code sequence. Consider $\rvec{x} = \rvec{v} + \rvec{u}_1 + \rvec{u}_2$, where $\rvec{v}$, $\rvec{u}_1$ and $\rvec{u}_1$ are three independent Gaussian codebook sequences. $\rvec{v}$ is an optimal Gaussian codebook sequence to $\rvec{z}$ of power $1-\beta$. $\rvec{u}_1$ is a secrecy capacity achieving Gaussian code sequence designed for reliable decoding at $\snr_u$ with the eavesdropper $\rvec{z}$ and power $\beta - \alpha$. Finally, $\rvec{u}_2$ is a secrecy capacity achieving code sequence for $\rvec{y}$ and eavesdropper $\rvec{z}$ and power $\alpha$. The decoding of each layer is possible by fully decoding the previous layers and considering any subsequent layers as additive Gaussian noise.
\end{IEEEproof}

\begin{IEEEproof}[Proof of Theorem \ref{thm:SNRu_low}]
First note that by complying with the constraint we have the following upper bound:
\begin{align}
\MMSE( \rvec{x} ; \gamma) \leq \frac{\alpha}{1 + \alpha \gamma}, \quad \forall \gamma \geq \snr_u.
\end{align}
Due to the decoding requirement at $\rvec{z}$ this translates also to an upper bound on $\MMSE( \rvec{x} ; \gamma | W_z)$
\begin{align}
\MMSE( \rvec{x} ; \gamma) = \MMSE( \rvec{x} ; \gamma | W_z) \leq \frac{\alpha}{1 + \alpha \gamma}, \quad \forall \gamma \geq \snr_z.
\end{align}
Denote
\begin{align} \label{eq:BCCdisturbance_proofAdditionalConstraint}
\MMSE( \rvec{x} ; \snr_z) = \MMSE( \rvec{x} ; \snr_z | W_z) = \frac{\beta}{1 + \beta \snr_z}.
\end{align}
If $\beta > \alpha$ we cannot comply with the constraint at $\snr_u$, and so we consider $\beta \in [0,\alpha]$. Due to the ``single crossing point'' property this choice will also comply with the upper bound for all $\gamma \geq \snr_z$. Moreover, this provides a lower bound on the following integral:
\begin{align}
\frac{1}{2} \int_0^{\snr_z} \MMSE(\rvec{x} ; \gamma | W_z) \d \gamma & \geq  \frac{1}{2} \int_0^{\snr_z} \frac{\beta}{1 + \beta \gamma} \d \gamma \nonumber \\
& = \frac{1}{2} \log ( 1 + \beta \snr_z).
\end{align}
Moreover, we have a direct upper bound on $\rate_y$, namely
\begin{align}
\rate_y & = \frac{1}{2} \int_{\snr_z}^{\snr_y} \MMSE( \rvec{x} ; \gamma) \d \gamma = \frac{1}{2} \int_{\snr_z}^{\snr_y} \MMSE( \rvec{x} ; \gamma | W_z) \d \gamma \nonumber \\
& \leq \frac{1}{2} \log \left( \frac{1 + \beta \snr_y}{1 + \beta \snr_z} \right).
\end{align}
It remains to upper bound $\rate_z$ under the constraint
\begin{align}
\rate_z & = \frac{1}{2} \int_0^{\snr_z} \left[ \MMSE(\rvec{x} ; \gamma) - \MMSE(\rvec{x} ; \gamma | W_z) \right] \d \gamma \nonumber \\
& \leq \frac{1}{2} \int_0^{\snr_z}  \MMSE(\rvec{x} ; \gamma) \d \gamma - \frac{1}{2} \log ( 1 + \beta \snr_z)
\end{align}
which is equivalent to maximizing the mutual information $\Igen{\rvec{x}}{\rvec{z}}$ subject to an MMSE constraint at some lower SNR. Note that we also need to consider the additional constraint we set at (\ref{eq:BCCdisturbance_proofAdditionalConstraint}); however we will disregard this constraint and see that due to the choice of $\beta \in [0, \alpha]$ we get this constraint ``for free''. Thus, the remaining problem is exactly the problem investigated in \cite{IMMSEtradeoff} and the solution is
\begin{align}
\rate_z & \leq \frac{1}{2} \log( 1 + \snr_u) + \frac{1}{2} \log\left( \frac{1 + \alpha \snr_z}{1 + \alpha \snr_u} \right) - \frac{1}{2} \log ( 1 + \beta \snr_z) \nonumber \\
& = \frac{1}{2} \log \left( 1 + \frac{(1- \alpha) \snr_u}{1 + \alpha \snr_u}\right) + \frac{1}{2} \log \left( 1 + \frac{(\alpha - \beta) \snr_z}{ 1 + \beta \snr_z}\right).
\end{align}
The above rate pair can be obtained by a 3-layer superposition code sequence. The first layer, of power $1- \alpha$, is an optimal Gaussian code sequence that can be reliably decoded at $\snr_u$ while considering all subsequent layers as additive Gaussian noise. The second layer is of power $\alpha - \beta$ and can be reliably decoded by $\rvec{z}$ after reliably decoding the first layer and considering the subsequent layer as additive Gaussian noise. The last layer is a secrecy capacity achieving code sequence of power $\beta$ designed for receiver $\rvec{y}$ with eavesdropper $\rvec{z}$.
\end{IEEEproof}

\section{Summary and Conclusions} \label{sec:Summary}
In this work we have considered several mutli-user scalar Gaussian settings in the unified framework of the scalar Gaussian channel.
Under this framework the Gaussian BC, the Gaussian wiretap channel, the Gaussian BCC and the Gaussian BC and Gaussian BCC with MMSE disturbance constraints have been defined using the requirements at different points on the SNR axis. Using the I-MMSE relationship we have proved MMSE properties required from any code sequence complying with these requirements. In some cases these properties completely define the specific family of code sequences.

More specifically, for the Gaussian wiretap channel, where apart from a reliable decoding requirement we also have a confidentiality requirement, we have shown that some well known ``rules of thumb'' used in achievability schemes are both necessary and sufficient for any ``good'' code sequence. The first being that, in any $d_{opt}$ code sequence, given the secret message the eavesdropper can fully decode the transmitted codeword. The second is that any $d_{\max}$ code sequence is built on a point-to-point achieving codebook sequence and that any partitioning of this codebook sequence to bins defines a completely secure message if and only if the bins themselves are codebook sequence that achieve capacity to the eavesdropper, that is, saturating the eavesdropper. It is also worth emphasizing that the rate-equivocation pair $(C, d_{\max})$ is an achievable pair for the Gaussian wiretap channel \cite{WynerWireTap,WireTapGaussian}, meaning that point-to-point capacity does not need to be compromised for the sake of maximum confidentiality, as shown in \cite[Ex. c, pp. 413]{CsiszarKorner} where the message is split to a completely secure message and an additional private message to obtain the entire rate-equivocation region.

The Gaussian BC has a unique and well-defined MMSE behavior for any ``good'' code sequence. This behavior holds regardless of the implementation, meaning the same behavior for both a superposition code sequence and a ``dirty paper'' code sequence. Not surprisingly, ``bad'' code sequences, that do not achieve capacity, are less defined, but a necessary and sufficient condition for the reliable decoding of the message to the weaker receiver requires that the conditioned MMSE (conditioned on that message) equals the MMSE of the codeword from the SNR of reliable decoding and onwards. This is not a surprising result but is worth mentioning due to the fact that strict concavity does not hold here, as opposed to any finite $\dim$ \cite[Theorem 2]{FunctionalPropertiesMMSE}. Finally, we derive necessary and sufficient conditions for a ``good'' code sequence, in terms of the MMSE and conditional MMSE behavior at all SNRs.

The Gaussian wiretap channel and the Gaussian BC come together to define the Gaussian BCC. In this setting we have examined two cases: complete secrecy and optimally secure code sequences. We first observed that ``good'' code sequences under both requirements have the same codeword MMSE as the Gaussian BC. The difference between the cases and compared with ``good'' sequences for the Gaussian BC is the behavior of the conditional MMSE quantities. For both problems we have provided an I-MMSE based converse proof to the capacity region.

Finally, we have considered also the effect of additional MMSE constraints at some other SNR. This continues the work in \cite{IMMSEtradeoff} where such constraints were originally examined, and provided interesting insights to the two-user Gaussian interference channel problem. Given the MMSE constraint we have examined both the Gaussian BC and the Gaussian BCC. In both cases it was necessary to distinguish between two possible regions for the SNR of the MMSE constraint. We have provided a capacity region proof for both problems and in the BCC case also a full depiction of the behavior of the different MMSE quantities.  We have observed that as in \cite{IMMSEtradeoff} the capacity achieving approach is superposition, however here we sometimes require a 3-layer superposition code sequence. Moreover, in the Gaussian BCC, depending on the SNR of the constraint, the second layer is either an optimal Gaussian code sequence to the eavesdropper (when the SNR of the constraint is lower than the SNR of the eavesdropper) or a Gaussian secrecy capacity achieving code sequence designed for a legitimate receiver at the constraint SNR and the eavesdropper (when the SNR is between the SNR of the eavesdropper and the legitimate receiver).

\appendix
\subsection{Applying Fatou's Lemma} \label{appendix:Fatou}
In this section we formally prove (\ref{eq:ReverseFatou}).
Examine the following sequence:
\begin{align}
\big| \MSEcoden(\gamma) - \MSEcode(\gamma)_{sup} \big|.
\end{align}
Note that by definition of the $\limsup$ (equivalent to (\ref{eq:defineMMSEsup})) we have that
\begin{align}
\limsup_{\dim \to \infty} \big| \MSEcoden(\gamma) - \MSEcode(\gamma)_{sup} \big| = 0.
\end{align}
Using the reverse Fatou's Lemma \cite{Lebesgue} (the MMSE function is bounded for all $\dim$) we have that
\begin{multline}
\limsup_{\dim \to \infty} \int \big| \MSEcoden(\gamma) - \MSEcode(\gamma)_{sup} \big| \d \gamma \leq \\ \int \limsup_{\dim \to \infty} \big| \MSEcoden(\gamma) - \MSEcode(\gamma)_{sup} \big| = 0.
\end{multline}
Thus, due to the non-negativity of the integrand we can conclude that
\begin{align} \label{eq:outcomeOfFatou}
\limsup_{\dim \to \infty} \int \MSEcoden(\gamma) \d \gamma = \int \MSEcode(\gamma)_{sup} \d \gamma.
\end{align}
On the other hand we assume that all normalized information quantities are stable ($\ie$, converge) meaning
\begin{align}
\lim_{\dim \to \infty} \frac{1}{\dim} \Igen{\rvec{x}_{\dim}}{\sqrt{\snr} \rvec{x}_{\dim} + \rvec{n}_{\dim}} = \limsup_{\dim \to \infty} \frac{1}{\dim} \Igen{\rvec{x}_{\dim}}{\sqrt{\snr} \rvec{x}_{\dim} + \rvec{n}_{\dim}}.
\end{align}
Due to the I-MMSE relationship \cite{IMMSE} we have that
\begin{align} \label{eq:outcomeIMMSE}
\lim_{\dim \to \infty} \int_0^{\snr} \MSEcoden(\gamma) \d \gamma = \limsup_{\dim \to \infty} \int_0^{\snr} \MSEcoden(\gamma) \d \gamma.
\end{align}
Putting (\ref{eq:outcomeOfFatou}) and (\ref{eq:outcomeIMMSE}) together we have that
\begin{align}
\lim_{\dim \to \infty} \int_0^{\snr} \MSEcoden(\gamma) \d \gamma = \int_0^{\snr} \MSEcode(\gamma)_{sup} \d \gamma.
\end{align}

\subsection{Proof of Theorem \ref{thm:ScalarUniqueCrossingPoint}} \label{appendix:SingleCrossingPointExt}
\begin{IEEEproof}
We define
\begin{align}
\CMSEr{\rvec{x}_{\dim,u}, \gamma}{\rvecr{y}_{\dim} } = \Esp[1]{ (\rvec{x}_{\dim,u} - \CEsp{\rvec{x}_{\dim,u}}{\rvecr{y}_{\dim}})
(\rvec{x}_{\dim,u} - \CEsp{\rvec{x}_{\dim,u}}{\rvecr{y}_{\dim}})^\T | \rvecr{y}_{\dim}}
\end{align}
Note that
\begin{align}
\MSE{\rvec{x}_{\dim}}(\scalart, u ) = \Esp{ \CMSEr{\rvec{x}_{\dim,u}, \gamma}{\rvec{y}_{\dim} } }
\end{align}
where the expectation is over $\rvec{y}_{\dim}$.
The following proof is similar in nature to the extension to the conditioned case discussed in \cite{Full_BustinPayaroPalomarShamai}.
It follows the proof of \cite[Theorem 1]{Full_BustinPayaroPalomarShamai}. We only emphasize the extension required for the conditioned case.
In \cite[Lemma 3]{Full_BustinPayaroPalomarShamai} it was shown that
\begin{align}
\Jacob_{\gamma} \q(\rvec{x}_{\dim} | \resc{u}=u, \sigma^2, \scalart) =  \frac{1}{\dim}\Tr \left( \Esp{\CMSEr{\rvec{x}_{\dim,{u}}, \gamma}{\rvec{y}_{\dim} }^2} \right)  - \frac{ \sigma^4}{(1 + \gamma \sigma^2)^2}.
\end{align}
Taking the expectation over $\resc{u}$ and using the linearity of the expectation and the trace function we have that
\begin{align}
\Jacob_{\gamma} \q(\rvec{x}_{\dim} | \resc{u}, \sigma^2, \scalart) =  \frac{1}{\dim}\Tr \left( \Esp{\CMSEr{\rvec{x}_{\dim,\resc{u}}, \gamma}{\rvec{y}_{\dim} }^2} \right)  - \frac{ \sigma^4}{(1 + \gamma \sigma^2)^2}.
\end{align}
We can now follow the derivation done in \cite[equations (20)-(24)]{Full_BustinPayaroPalomarShamai} and conclude that for all $\q(\rvec{x}_{\dim} | \resc{u}, \sigma^2, \scalart) < 0$ the derivative with respect to $\gamma$ is positive. Following the proof of \cite[Theorem 1]{Full_BustinPayaroPalomarShamai} we conclude the proof.
\end{IEEEproof}

\subsection{Theorem \ref{thm:ScalarUniqueCrossingPoint} as $\dim \to \infty$} \label{appendix:singleCrossing}
In this section we show that the ``single crossing point'' property given in Theorem \ref{thm:ScalarUniqueCrossingPoint} holds also for
\begin{align}
\liminf_{\dim \to \infty} \q( \rvec{x}_{\dim} | \resc{u}, \sigma^2, \gamma) = \lim_{\dim \to \infty} \left( \inf_{m \geq \dim} \q( \rvec{x}_{m}| \resc{u}, \sigma^2, \gamma) \right).
\end{align}
As such we can conclude a ``single crossing point'' property between the $\iid $ Gaussian input and the $\limsup \MMSE$ function, whenever the limit of the MMSE function does not exist (and with the $\lim \MMSE$ when it does exist). For this purpose we show that, for any natural $\dim$, the function
\begin{align}
\inf_{m \geq \dim} \q( \rvec{x}_{m}| \resc{u}, \sigma^2, \gamma)
\end{align}
has at most a single crossing point. Since this function is monotonically increasing in $\dim$ (and bounded, so that it has a limit), the fact that it has at most a single crossing point suffices to conclude at most a single crossing point for the limit. Recall that for every natural $\dim$ $\q( \rvec{x}_{\dim}| \resc{u}, \sigma^2, \gamma)$ has at most a single crossing point, meaning that if for some $\gamma$
\begin{align}
\inf_{m \geq \dim} \q( \rvec{x}_{\dim}| \resc{u}, \sigma^2, \gamma) \geq 0
\end{align}
the crossing point has occurred for all $\q( \rvec{x}_{m}| \resc{u}, \sigma^2, \gamma)$, $m \geq \dim$, and we can conclude that
\begin{align}
\inf_{m \geq \dim} \q( \rvec{x}_{\dim}| \resc{u}, \sigma^2, \gamma') \geq 0, \quad \forall \gamma' \geq \gamma,
\end{align}
meaning that no transition from a non-negative value to a negative value can occur; hence there is at most a single crossing point.




\bibliographystyle{IEEEtran}

\bibliography{bib}

\end{document}